\documentclass[12 pt]{article}

\usepackage[a4paper,bindingoffset=0.2in,%
            left=1in,right=1in,top=1in,bottom=1in,%
            footskip=.25in]{geometry}
\usepackage{authblk}
\usepackage{amssymb}
\usepackage[figuresright]{rotating}
\usepackage[T1]{fontenc}
\usepackage{amsfonts}
\usepackage{dcolumn}
\usepackage{amssymb}
\usepackage{amsmath}
\usepackage{graphicx}
\usepackage{supertabular}
\usepackage{bm}
\usepackage{dsfont}
\usepackage{braket}
\usepackage{hyperref}
\usepackage{color,soul}
\usepackage[english]{babel}

\begin{document}

\title{Lattice dynamics and thermophysical properties of h.c.p. Os and Ru from the quasi-harmonic approximation}

\author{Mauro Palumbo [1], Andrea Dal Corso [1,2]}
\affil{[1] International School for Advanced Studies (SISSA), Trieste, ITALY}
\affil{[2] CNR-IOM, Trieste, ITALY}

\maketitle   


\begin{abstract}
  We report first-principles phonon frequencies and anharmonic thermodynamic properties of h.c.p. Os and Ru calculated 
  within the quasi-harmonic approximation, including Gr\"uneisen parameters, temperature-dependent lattice parameters, thermal expansion, 
  and isobaric heat capacity.
  We discuss 
  the differences between a full treatment of anisotropy and a simplified approach with a constant $c/a$
  ratio. 
  The results are systematically compared with the available experimental data and an overall satisfactory agreement is obtained. 
\end{abstract}

\newpage
\section{Introduction}

Osmium (Os) and Ruthenium (Ru), both belonging to the same group (VIII B) of the periodic table, are hexagonal elements with high melting points, densities, and cohesive energies.
The former has the highest measured density among the elements~\cite{14Arb,91You}, it is very hard and brittle even at high temperatures, and it is used in Pt and Ir alloys
to enhance their mechanical properties and in the chemical industry as osmium tetroxide. 
Ruthenium finds use in electronic industry for manufacturing electrical contacts
and chip resistors as well as in chemical industry for use as anodes for chlorine production in electrochemical cells.

The experimental determination of phonon dispersions and thermophysical properties of both elements is difficult.  
Only a few phonon frequencies have been measured for Os using Raman spectroscopy~\cite{05Pon,98Pon}, 
while for Ru phonon dispersions have been obtained recently
using inelastic neutron scattering~\cite{00Hei}. A previous work also reports experimental phonon frequencies along the [001]
direction ~\cite{81Smi}. Available crystallographic data on Os have been reviewed by Arblaster~\cite{13Arb}, who 
reports assessed values of the lattice parameters and thermal expansion coefficients from 0 to 1300~K and references to prior experimental 
work~\cite{37Owe,71Fin,72Sch}. The available thermodynamic properties of Ru are reported in~\cite{95Arb} and references therein. 
Arblaster~\cite{95Arb, 95Arb2} also reports assessed values on thermodynamic quantities and crystallographic properties of Ru. 
Additional experimental work can be found in Refs.~\cite{37Owe,57Hal,62Ros,71Fin,72Sch}.

Several theoretical works have been dedicated to h.c.p. Os~\cite{16Tal,15Dub,11Den,11Liu,05Ma}. 
An experimental investigation using the diamond anvil 
cell~\cite{02Cyn} found that the bulk modulus of Os was 462 GPa, higher than diamond, but later studies reported lower values~\cite{04Tak,04Occ}.
The most recent theoretical studies have focused 
on the high-pressure behavior of this element where an anomaly in the first derivative of the $c/a$ ratio with respect 
to pressure has been found 
experimentally at 25~GPa~\cite{04Occ}. This anomaly was later disproven by Dubrovinsky et al.~\cite{15Dub},
who report a detailed experimental and theoretical investigation of Os behavior at very high pressure (up to 770 GPa)
using state-of-the-art methods, including dynamical mean field theory (DMFT) and spin-orbit coupling, but without computing phonons. They found anomalies at approximately 150 GPa 
and 440 GPa.
Temperature-dependent properties have been calculated by Liu et al.~\cite{11Liu} and Deng et al.~\cite{11Den}. The former used a quasi-harmonic Debye model,
where the temperature-dependence of the free energy is obtained from a volume-dependent Debye temperature and not from phonon frequencies. On the contrary,
Deng et al. performed phonon calculations for different volumes and then applied a Birch equation of state (EOS) to derive quasi-harmonic quantities.
The phonon spectra of Ru has been theoretically investigated by Heid et. al.~\cite{00Hei} and Souvatzis et al.~\cite{08Sou}. The latter authors 
have also applied a quasi-harmonic approach, including the evaluation of the thermal expansion, similarly to Yaozhuang et al.~\cite{07Yao}. 
We remark that anomalies found by Dubrovinsky et al. occur at pressures well above the range of quasi-harmonic calculations. 

With the exception of the work by Souvatzis et al., however, all previous theoretical quasi-harmonic results rely on ultrasoft pseudopotentials 
or earlier-developed methods.
Published works reporting results at the quasi-harmonic level use only one of the well-established LDA/PBE functionals without an
extensive comparison with other functionals, previous theoretical results and experimental data. Besides, some properties such as mode Gr\"uneisen
parameters and the temperature dependence of both $a$ and $c$ lattice parameters were not reported in the literature and anisotropic linear thermal
expansions were reported only for Ru. 
 
In this paper we report anharmonic results calculated with the quasi-harmonic approximation using PAW pseudopotentials from the 
Quantum Espresso pslibrary1.0~\cite{pslibrary,14Dal}. We test three different functionals (LDA, PBE and PBEsol) at the same level of theory.
We extensively compare our results with an extended set of experimental data and previous theoretical results aiming at providing a consistent
set of reliable theoretical data within the quasi-harmonic approximation. Finally we investigate the possibility that also Os and Ru have anomalous features of the phonon dispersions
as those we found recently for Re and Tc~\cite{paperReTc}, i.e. the presence of soft modes at the $\Gamma$ and H points in the Brillouin zone.
Anomalies were also found in the Gr\"uneisen parameters of Re and Tc and are studied here
for Os and Ru.

Phonon dispersions were computed for several volumes and lattice parameters. From the phonon frequencies, the Gr\"uneisen parameters, the thermal expansion 
and other thermophysical quantities as a function of temperature were derived. The results were first calculated
assuming a constant $c/a$ ratio and computing volume dependent phonon frequencies. A full treatment of anisotropy in the hexagonal cell was then applied
to obtain the temperature dependence of $a$ and $c$ and the thermal expansion tensor.

\section{Computational method}\label{sec:Compmethod}

All calculations in the present work were carried out using Density Functional Theory (DFT) and plane waves basis sets as implemented in the 
Quantum Espresso (QE) package~\cite{QEweb,09Gia}. The thermo\_pw package~\cite{thermopw} was used for obtaining harmonic and quasi-harmonic properties.
The LDA~\cite{81Per}, the PBE-GGA~\cite{96Per} and the revised PBE functional for densely-packed solids (PBEsol)~\cite{08Per} 
were tested for the exchange-correlation functional. We employed the
Os.pz-spn-kjpaw\_psl.1.0.0.UPF, Os.pbe-spn-kjpaw\_psl.1.0.0.UPF, Os.pbesol-spn-kjpaw\_psl.1.0.0.UPF, Ru.pz-spn-kjpaw\_psl.1.0.0.UPF, Ru.pbe-spn-kjpaw\_psl.1.0.0.UPF
and Ru.pbesol-spn-kjpaw\_psl.1.0.0.UPF pseudopotentials from the pslibrary 1.0~\cite{pslibrary,14Dal}.

We have chosen computational settings to ensure that all investigated properties and in particular phonon frequencies are well converged. 
The kinetic energy cut-off was set to 50~Ry while the charge density cut-off was set to 200~Ry, except for Os PBE and PBEsol for which it was set to 300~Ry. 
The integration over the Brillouin Zone (BZ)
was  performed employing a 18$\times$18$\times$12 Monkhorst-Pack $\vec{k}$-points mesh 
and a Methfessel-Paxton smearing scheme with 0.02~Ry width.
The Density Functional Perturbation Theory (DFPT) as implemented in the phonon code~\cite{01Bar,08Dal} 
was used to calculate phonon frequencies on a 6$\times$6$\times$4 grid of
$\vec{q}$-points and a Fourier interpolation was used for other points in the BZ.

In the quasi-harmonic approximation~\cite{10Bar}, the Helmholtz energy of a crystalline solid is 
\begin{equation}
F(T,X) = U_0(X)+F^{\rm{vib}}(T,X)+F^{\rm{el}}(T,X)
\label{eq:F}
\end{equation}
where $U_0$ is the static energy at 0~K, $F^{\rm{vib}}$ the contribution due to lattice vibrations and $F^{\rm{el}}$ the energy due to electronic
excitations.  In the adiabatic approximation, each term is treated separately.
X refers to any variable upon which the above energies may depend and most often it is the unit cell volume as for example for cubic lattices. 
In the case of an h.c.p.
cell, X refers to the two lattice parameters $a$ and $c/a$, although it is common to assume that $c/a$ is constant
as a function of temperature and thus consider the variation of $F$ only as a function of $a$ (or equivalently $V$). 
For a given $X$, the vibrational Helmholtz energy per cell is calculated as in the harmonic approximation: 
\begin{equation}
 F^{\rm{vib}}(X,T) =\frac{1}{2 N} \sum_{\vec{q},\nu}{\hbar \omega(\vec{q},\nu,X)} + \\
 \frac{k_{\rm B} T}{N} \sum_{\vec{q},\nu}\ln[1-\exp(\frac{-\hbar \omega(\vec{q},\nu,X)}{k_{\rm B} T})].
\label{eq:Fharm}
\end{equation}
The first term on the right-hand side of the above equation is the Zero-Point Energy (ZPE), the second term is the phonon contribution at finite
temperatures. The sums in Eq.~(\ref{eq:Fharm}) are taken over the phonon frequencies $\omega(\vec{q},\nu,X)$, where $\nu$ denotes the different
phonon branches and $\vec{q}$ the wave vectors within the first BZ. $k_B$ is the Boltzmann constant, $\hbar$ the reduced Planck constant, 
$T$ the absolute temperature, and $N$ the number of unit cells in the solid.

We carried out calculations both (a) assuming a constant $c/a$ and (b) evaluating the Helmholtz energy on a full grid ($a$, $c/a$).

In the former case, we computed the phonon frequencies at 9 different volumes (or equivalently $a$ values), keeping the $c/a$ 
constant at the 0~K 
equilibrium value and fitting the resulting energies using a Murnaghan EOS~\cite{44Mur}. 
The minimum $U_0(V)+F^{\rm{vib}}(T,V)$ at each temperature corresponds to zero pressure. In this way, the temperature dependence
of the free energy $F$, the volume $V$ and the isothermal bulk modulus $B_T$ is directly obtained from the Murnaghan EOS. 
By numerical derivation 
of the volume we obtained the volume thermal expansion coefficient $\beta$ according to its definition
\begin{equation}
 \beta =\frac{1}{V} \left(\frac{\partial V}{\partial T}\right)_{P}.
\label{eq:beta}
\end{equation}
The so-called Gr\"{u}neisein mode parameters were calculated as
\begin{equation}
\gamma_{{\bf q},\nu} = - {V\over \omega(\vec{q},\nu,V)} {\partial
\omega(\vec{q},\nu,V) \over \partial V} ,
\label{eq:Grun}
\end{equation}
from the computed frequencies. 
The constant volume (isochoric) heat capacity was obtained from the phonon frequencies calculated as
in the harmonic approximation
\begin{equation}
 C_{V} =\frac{k_{\rm B}}{N} \sum_{\vec{q},\nu}\left(\frac{\hbar\omega(\vec{q},\nu)}{k_BT}\right)^2 \frac{\exp(\hbar\omega(\vec{q},\nu)/k_B T)}{\left[\exp(\hbar\omega(\vec{q},\nu)/k_B T)-1 \right]^2},
\label{eq:Cv}
\end{equation}
at each fixed volume in the quasi-harmonic grid. The isochoric heat capacity at each temperature is then obtained 
interpolating at the temperature-dependent volume values obtained at each temperature from the minimization of the free energy.
Finally, the constant pressure (isobaric) heat capacity was obtained as~\cite{72Wal}
\begin{equation}
 C_{P} = C_{V} + TV\beta^{2}B_{T} .
\label{eq:Cp}
\end{equation}

In the full anisotropic case (b), the calculations
were carried out on a grid of points of $a$ and $c/a$ and then the Helmholtz energy at each T was fitted with quartic polynomials as a function of $a$ and $c/a$. 
For both Os and Ru, a total of 25 grid points ($5\times5$)
were used with steps of 0.05~a.u. and 0.02 in $a$ and $c/a$, respectively. The grid was approximately centered 
at the equilibrium values of $a$ and $c/a$ at 0~K. We derive the variation with temperature of $a$ and $c$ from the minimization of $F$.

The components of the thermal expansion are obtained from the temperature variation of the lattice parameters as

\begin{equation}
\alpha_1= \alpha_2 = {1\over a(T) } { d a(T) \over d T },
\label{alphaa}
\end{equation}

\begin{equation}
\alpha_3= {1\over c(T) } {d c(T) \over d T},
\label{alphac}
\end{equation}
since in the present hexagonal case only two thermal expansion terms are independent. 
Finally, the volume thermal expansion $\beta$ is given by $\beta=2 \alpha_1 + \alpha_3$.

The electronic contribution to the heat capacity was obtained in the single-particle approximation from the electronic DOS 
at the equilibrium lattice parameters at 0~K, as described in details in Ref.~\cite{14Pal,99Gri}.

More details on our approach are reported in Ref.~\cite{paperReTc}.

\section{Results}

\subsection{Osmium}

The calculated lattice parameters of h.c.p. Os are first compared with experimental and previous theoretical results from different sources 
in Tab.~\ref{tab:Oslatticeparameters}.
The experimental data have been taken from Ref.~\cite{13Arb}, where the lattice parameters values have been corrected to 293~K using thermal expansion data.
It can be noted that the PBEsol results are in remarkable good agreement with the experiments. There is also a good agreement with 
previous theoretical results at 0~K, despite the fact that some of the earlier works were using different methods, with the only exception of the early
work by Fast et al. whose LDA values appear to be higher than all other LDA results.

\begin{table}[htp]
\caption{\label{tab:Oslatticeparameters} Comparison between structural parameters of h.c.p. Os from different sources.  
Experimental data are taken from Ref.~\cite{13Arb} and references therein. They have been corrected to 293 K using thermal expansion data.
Previous theoretical results are also reported for comparison (Liu et al. 2011~\cite{11Liu}, Ma et al. 2005~\cite{05Ma}, Sahu et al. 2005~\cite{05Sah}, 
Deng et al. 2009~\cite{09Den}, Cynn et al. 2002~\cite{02Cyn}, Fast et al. 1995~\cite{95Fas}). 
US is for ultrasoft pseudopotentials, PAW for projector-augmented-wave pseudopotentials , FPLMTO for full-potential linear muffin-tin orbitals,
FPLAP for full-potential linearized augmented plane-waves.
}
\begin{tabular}{@{}lcccc}
 Method & $a$ & $c$ & $c/a$ \\
        & $\AA$ & $\AA$ &  \\
PAW, LDA (0~K, this work) & 2.7164 & 4.2879 & 1.5785  \\
PAW, PBEsol (0~K, this work) & 2.7284 & 4.3062 & 1.5783  \\
PAW, PBE (0~K, this work) & 2.7540 & 4.3441 & 1.5774  \\
PAW, LDA (293~K, this work) & 2.7200 & 4.2966 & 1.5796 \\
PAW, PBEsol (293~K, this work) & 2.7321 & 4.3154 & 1.5795  \\
PAW, PBE (293~K, this work) & 2.7579 & 4.3540 & 1.5787 \\
US, LDA (0~K, Liu et al. 2011) & 2.7135  & 4.2737  & 1.575  \\
US, PBE (0~K, Liu et al. 2011)  & 2.7489  & 4.3350  & 1.577  \\
US, PBE (0~K, Deng et al. 2009) & 2.745  & 4.328  & 1.577  \\
US, PBE (0~K, Ma et al. 2005) & 2.7507 & 4.3388 & 1.5773 \\
FPLAP, LDA (0~K, Sahu et al. 2005) & 2.7204 & 4.3112 & 1.5847 \\
FPLAP, PBE (0~K, Sahu et al. 2005) & 2.7576 & 4.3667 & 1.5835 \\
FPLMTO, LDA (0~K, Cynn et al. 2002) & 2.7165 & 4.3029 & 1.5840 \\ 
FPLMTO, LDA (0~K, Fast et al. 1995) & 2.752 & 4.344 & 1.578 \\
Exp. (Owen et al. 1935)    & 2.7361 & 4.3189 & 1.5785 \\
Exp. (Owen and Roberts 1936)    & 2.7357 & 4.3191 & 1.5788 \\
Exp. (Owen and Roberts 1937)    & 2.7355 & 4.3194 & 1.579 \\
Exp. (Finkel et al. 1971)    & 2.7346 & 4.3174 & 1.5788 \\
Exp. (Rudman 1965)    & 2.7341 & 4.3188 & 1.5796 \\
Exp. (Swanson et al. 1955)    & 2.7342 & 4.3198 & 1.5799 \\
Exp. (Mueller and Heaton 1961)    & 2.7345 & 4.3200 & 1.5798 \\
Exp. (Taylor et al. 1961)    & 2.7342 & 4.3201 & 1.5800 \\
Exp. (Schroeder et al. 1972)    & 2.7340 & 4.3198 & 1.5800 \\
Exp. (Liu et a. 2011)  &  2.7356  & 4.3198  &  1.579  \\
\end{tabular}
\end{table}

\begin{figure}[htp]
\includegraphics[height=10cm]{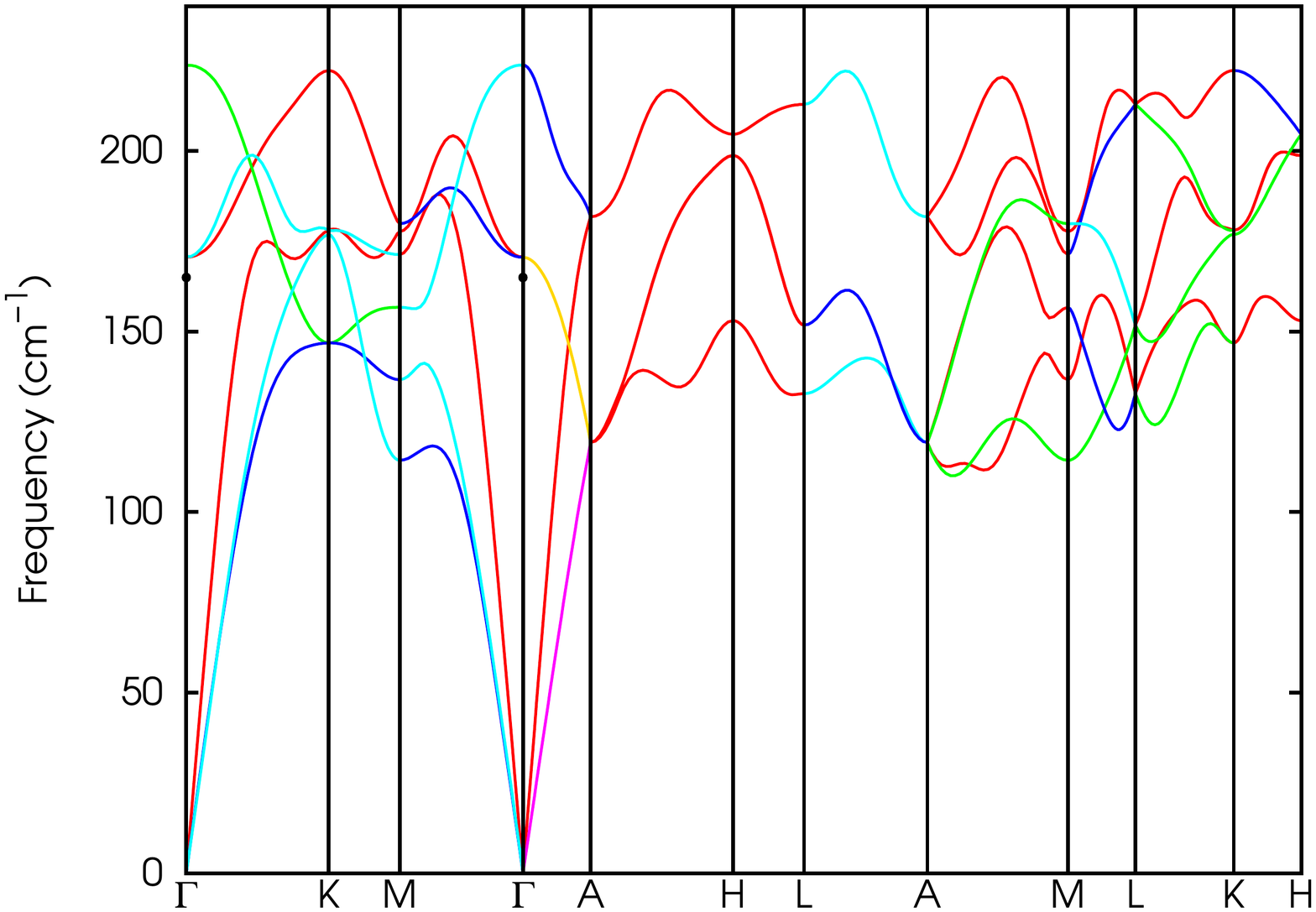}
\includegraphics[height=10cm]{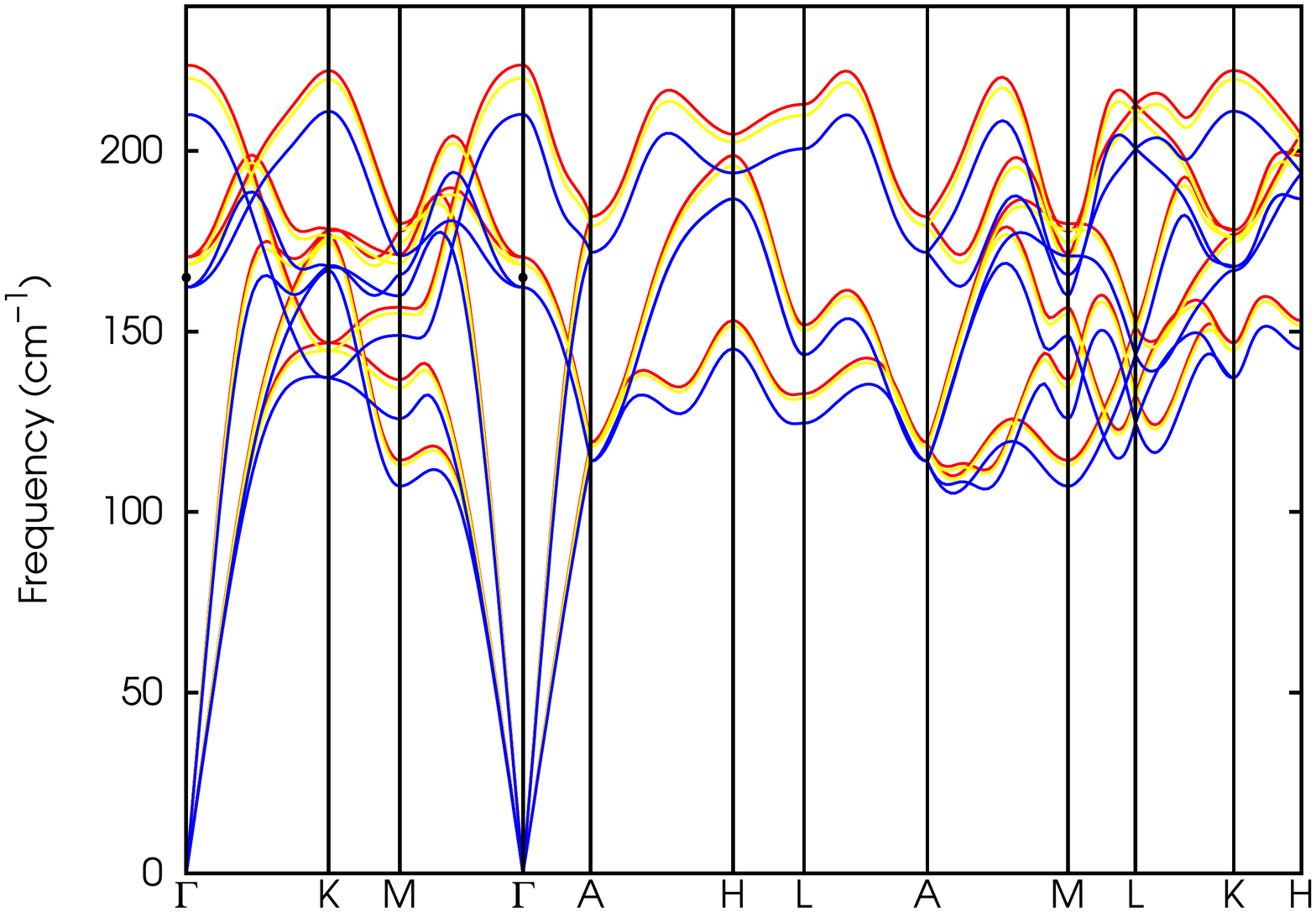}
\caption{(a) Phonon dispersions of h.c.p. Os calculated with
LDA exchange-correlation functional. The colours represent the different simmetries of the phonon branches (for details see the thermo\_pw
documentation); (b) Phonon dispersions of h.c.p. Os calculated with
LDA (red line), PBEsol (yellow line) and PBE (blue line) exchange-correlation functionals.
All calculations were carried out interpolating phonon dispersions at the equilibrium lattice parameters at 300~K.
}\label{fg:Osdisp}
\end{figure}

\begin{figure}[htp]
\includegraphics[height=10cm]{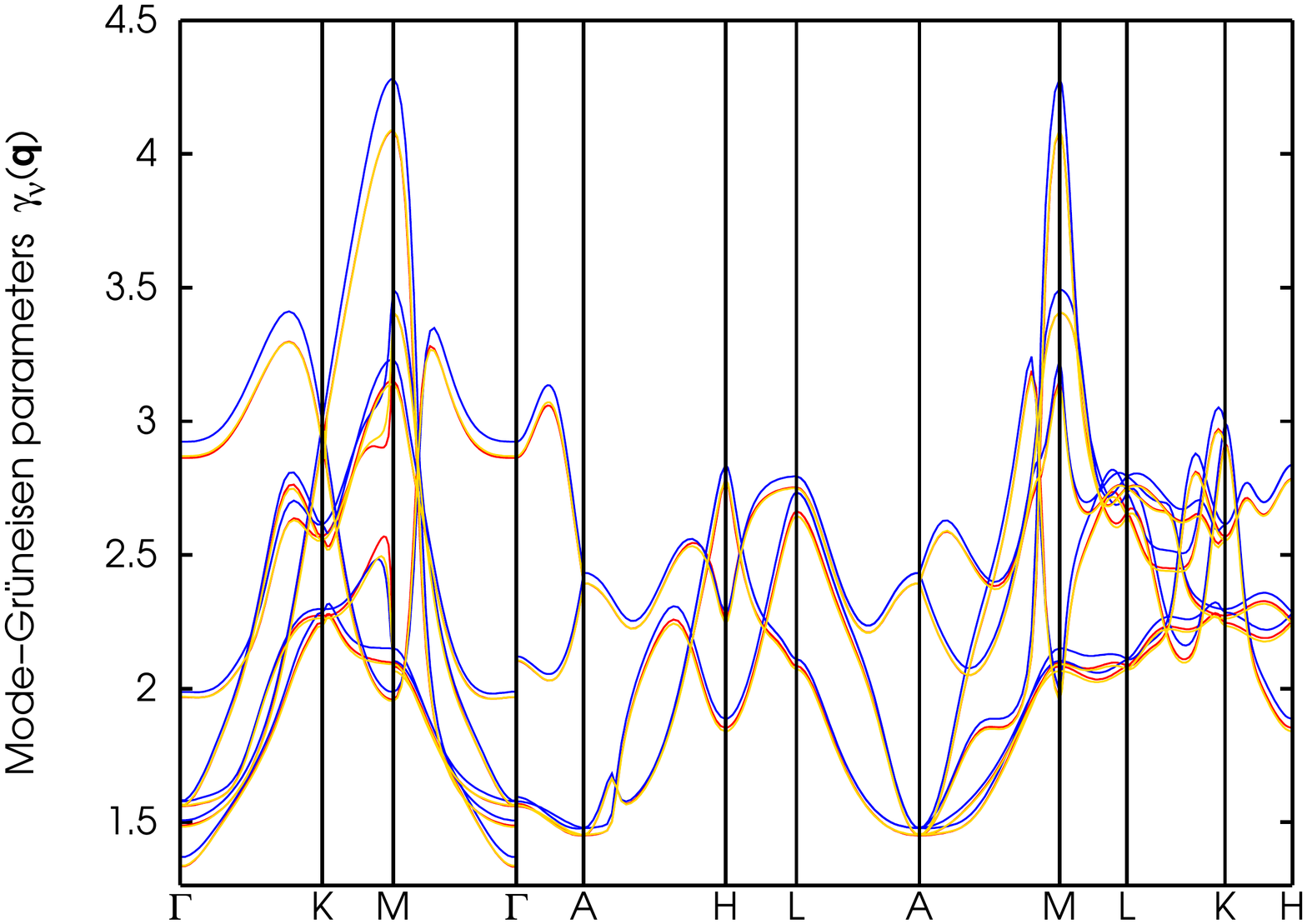}
\caption{Volume-dependent mode Gr\"{u}neisein parameters of h.c.p. Os calculated with LDA (red line), PBEsol (yellow line) and PBE (blue line)
exchange-correlation functionals at the equilibrium 
lattice parameters at 300~K.  
}\label{fg:Osgrun}
\end{figure}

Lattice dynamics results for h.c.p. Os are plotted in Fig.~\ref{fg:Osdisp} in the LDA approximation. The calculated phonon dispersion
is similar to that obtained by Deng et al.~\cite{11Den}. 
The only available experimental point at $\Gamma$ is also shown for comparison and is about 5 $cm^{-1}$ lower than the
calculated LDA value.
The phonon dispersion is also similar to that 
of other h.c.p. metals such as Re and Tc~\cite{paperReTc} which are close to Os in the periodic table. An important difference is, however,
the absence in Os of soft modes at
$\Gamma$ and H contrary to what has been found in both Re and Tc. At these same points, the calculated Gr\"uneisen parameters of Os (Fig.~\ref{fg:Osgrun})
show no anomalous peaks contrary to Re and Tc, hence the phonon frequencies do not rapidly become imaginary when increasing the volume. We note that no negative values
of the Gr\"uneisen parameters occur in agreement with positive values of the thermal expansion tensor at any temperature (see below).

\begin{figure}[htp]
\includegraphics[height=10cm]{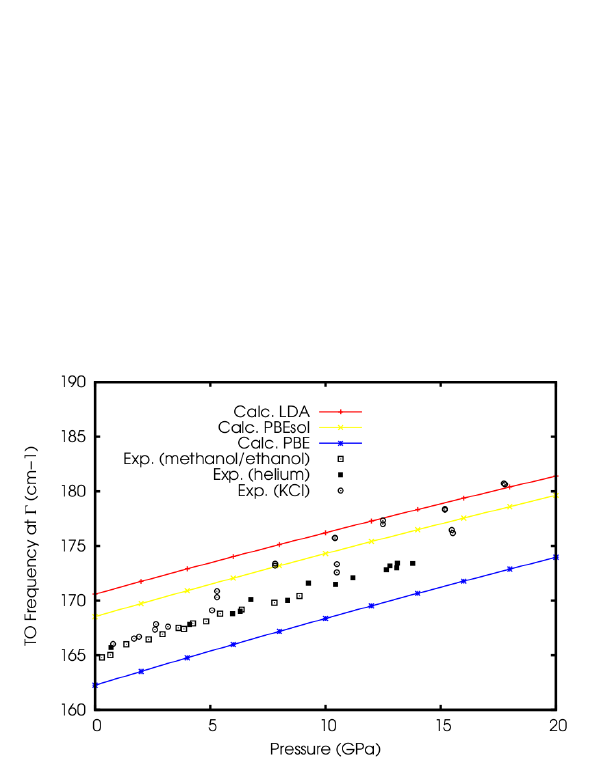}
\caption{Pressure shift of the TO phonon frequency at the $\Gamma$ point and T=300~K with LDA, PBEsol and PBE functionals.
Experimental data were obtained from Raman spectroscopy in Ref.~\cite{05Pon} and using three different pressure-transmitting 
media as in the legend.}\label{fg:freqvsP}
\end{figure}

The pressure shift of the TO mode measured at $\Gamma$ using Raman spectroscopy and three different pressure-transmitting media
(Ref.~\cite{05Pon}) are compared to our results in Fig.~\ref{fg:freqvsP}. The frequencies where calculated at 300~K and for each pressure value using
the equilibrium geometry obtained by minimizing the Gibbs energy. 
It can be noted that the experimental data are in between the LDA and PBE calculated values, with the PBEsol results in better agreement. The experimental
values show some scatter, in particular the data set obtained
using KCl as a pressure-transmitting medium are closer to our PBEsol results. The calculated 
temperature dependence of the TO mode is, however, similar to the experimental results obtained using helium and methanol/ethanol, which appear to be
consistent with each other. The values of the equilibrium lattice parameters $a$ and $c/a$ at 300~K as a function of pressure 
obtained from the minimization of the free energy
are reported in Tab.~\ref{tab:acaP}. The frequencies in Fig.~\ref{fg:freqvsP} were computed at these lattice parameters values 

\begin{table}[htp]
\caption{\label{tab:acaP} Structural parameters of h.c.p. Os at 300~K as a function of pressure according to different DFT functionals.  
The values reported here correspond to those for which the frequencies at the $\Gamma$ point in Fig.~\ref{fg:freqvsP} were obtained.}
\begin{tabular}{@{}ccccccc}
 Pressure    & $a$ (LDA) & $c/a$ (LDA) & $a$ (PBEsol) & $c/a$ (PBEsol) & $a$ (PBE) & $c/a$ (PBE) \\
   (GPa)     & $\AA$ &  & $\AA$ &  & $\AA$ &  \\
0  & 2.7201 & 1.5796  & 2.7321 & 1.5796 & 2.7580 & 1.5788  \\
2 & 2.7158 & 1.58 & 2.7277 & 1.5799 & 2.7531 &  1.5791  \\
4 & 2.7116 & 1.5803 & 2.7233 & 1.5802 & 2.7483 &  1.5795  \\
6 & 2.7075 & 1.5806 & 2.7191 & 1.5805 & 2.7436 &  1.5798  \\
8 & 2.7035 & 1.5809 & 2.7149 & 1.5808 & 2.7391 &  1.5801  \\
10 & 2.6996 & 1.5812 & 2.7108 & 1.5811 & 2.7346 &  1.5805  \\
12 & 2.6957 & 1.5815 & 2.7068 & 1.5814 & 2.7303 &  1.5808  \\ 
14 & 2.6920 & 1.5818 & 2.7029 & 1.5817 & 2.7261 &  1.5811  \\ 
16 & 2.6883 & 1.5821 & 2.6991 & 1.582 & 2.7219 &  1.5814  \\
18 & 2.6847 & 1.5823 & 2.6954 & 1.5823 & 2.7179 &  1.5816  \\
20 & 2.6811 & 1.5826 & 2.6917 & 1.5825 & 2.7139 &  1.5819  \\
\end{tabular}
\end{table}

\begin{figure}[htp]
\hbox{\includegraphics[height=6cm]{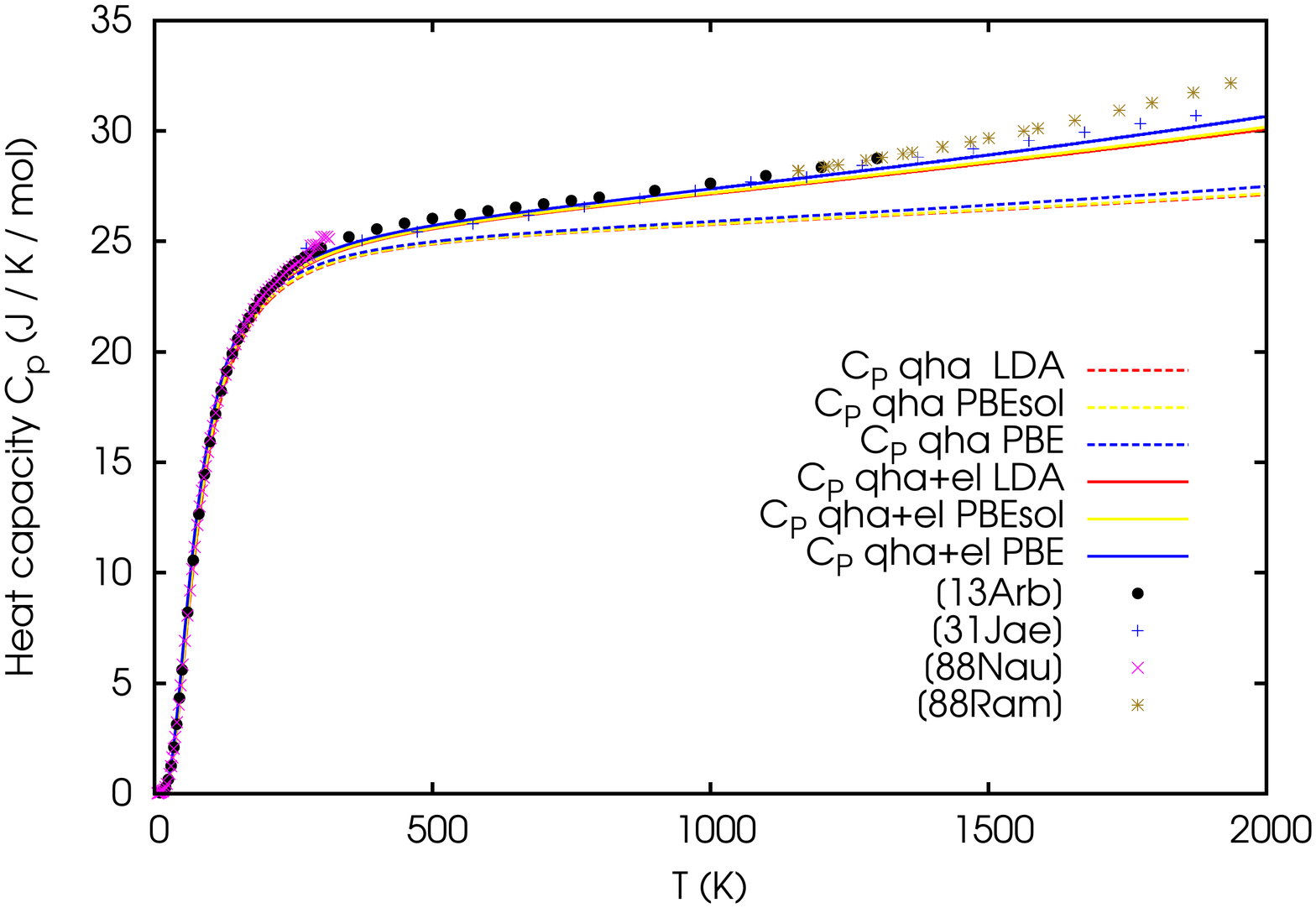}
\includegraphics[height=6cm]{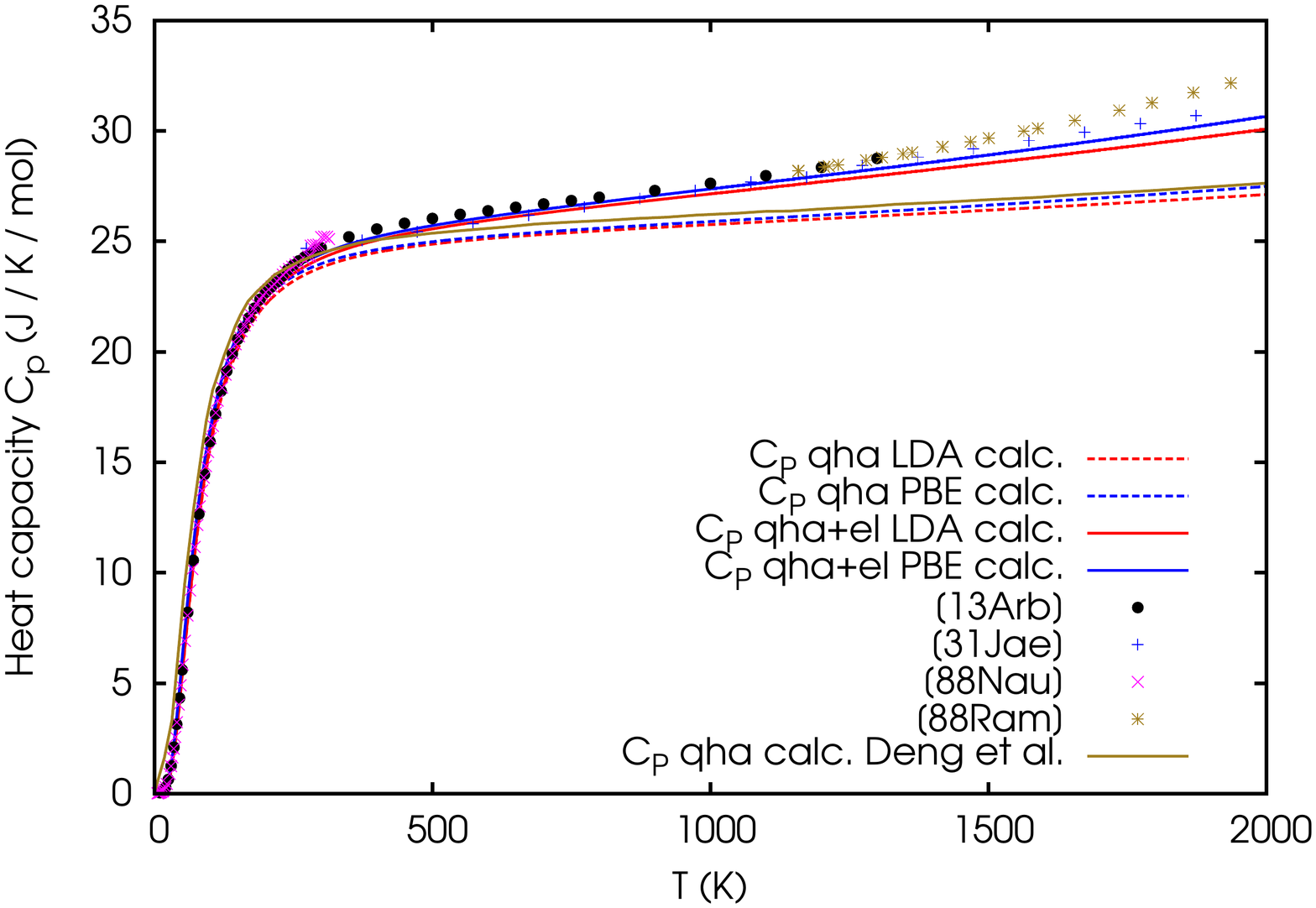}}
\caption{(a) Calculated isobaric heat capacity of h.c.p. Os as a function of temperature with LDA/PBEsol/PBE exchange-correlation functionals, 
quasi-harmonic only ($C_P$ qha) and 
quasi-harmonic plus electronic ($C_P$ qha+el). The calculations were done assuming a constant $c/a$ ratio and using the Murnaghan EOS. 
Experimental data from different sources are shown as points: 31Jae~\cite{31Jae}, 88Nau~\cite{88Nau} and 88Ram~\cite{88Ram}. Points from 91Arb (ref.~\cite{91Arb}) 
are not pure experimental data but assessed values obtained from a critical evaluation of available experimental data; (b) As in (a) but including
previously calculated data from Deng et al. (and removing PBEsol results for clarity).
}\label{fg:OsCp}
\end{figure}

As outlined in the previous section, we first performed quasi-harmonic calculations assuming a constant $c/a$ ratio. As an example of the results obtained in
this way, we show in Fig.~\ref{fg:OsCp} the isobaric heat capacity of h.c.p. Os calculated using both the LDA and PBE functionals. The figure also shows
available experimental data for comparison. The points from Arblaster~\cite{91Arb} are not pure experimental data but the result of a critical
evaluation of the available experiments up to 1300 K. It is noteworthy that the electronic contribution, as common for metals, is not negligible and is
essential to reach a satisfactory agreement with the experimental data. It can be remarked that the agreement is good at low temperature, while the calculated 
line departs from the experimental points at high temperatures. This is not unexpected, since the quasi-harmonic approximation breaks down when 
higher-order anharmonic contributions become important at high temperatures.
As mentioned in the introduction, theoretical calculations for thermodynamic properties of h.c.p. Os were reported 
by Liu et al.~\cite{11Liu} and Deng et al.~\cite{11Den}. The former however did not report any isobaric heat capacity results. The results of Deng et al.,
are shown in Fig.~\ref{fg:OsCp} (olive line) and they are very close to ours without including the electronic contribution.
It is noteworthy that they used a different EOS (Birch) and they computed
volume-dependent phonon frequencies relaxing the $c/a$ ratio at each fixed volume. They did not compute temperature-dependent lattice parameters and 
the linear thermal expansions.

\begin{figure}[htp]
\hbox{
\includegraphics[height=6.5cm]{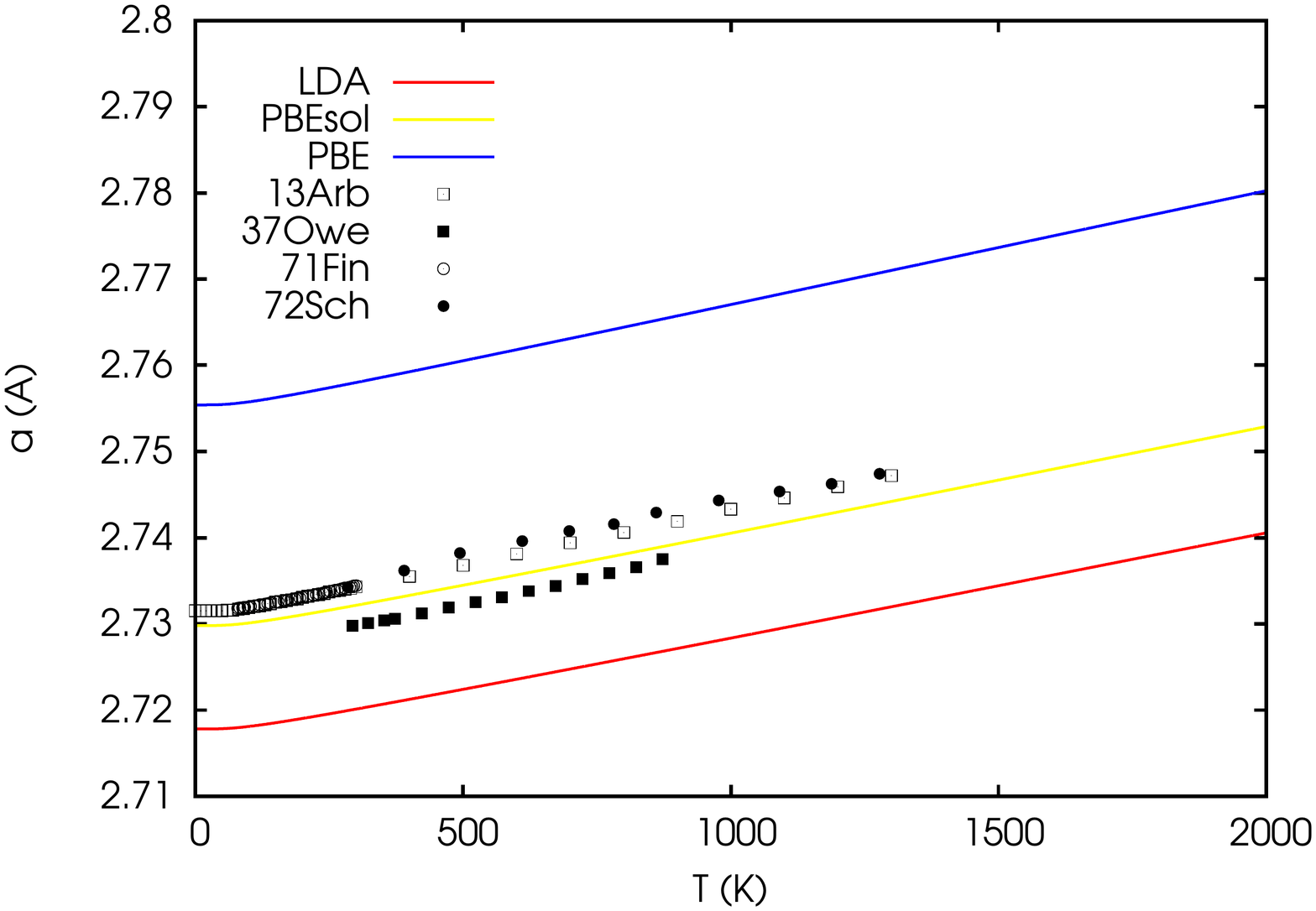}
\includegraphics[height=6.5cm]{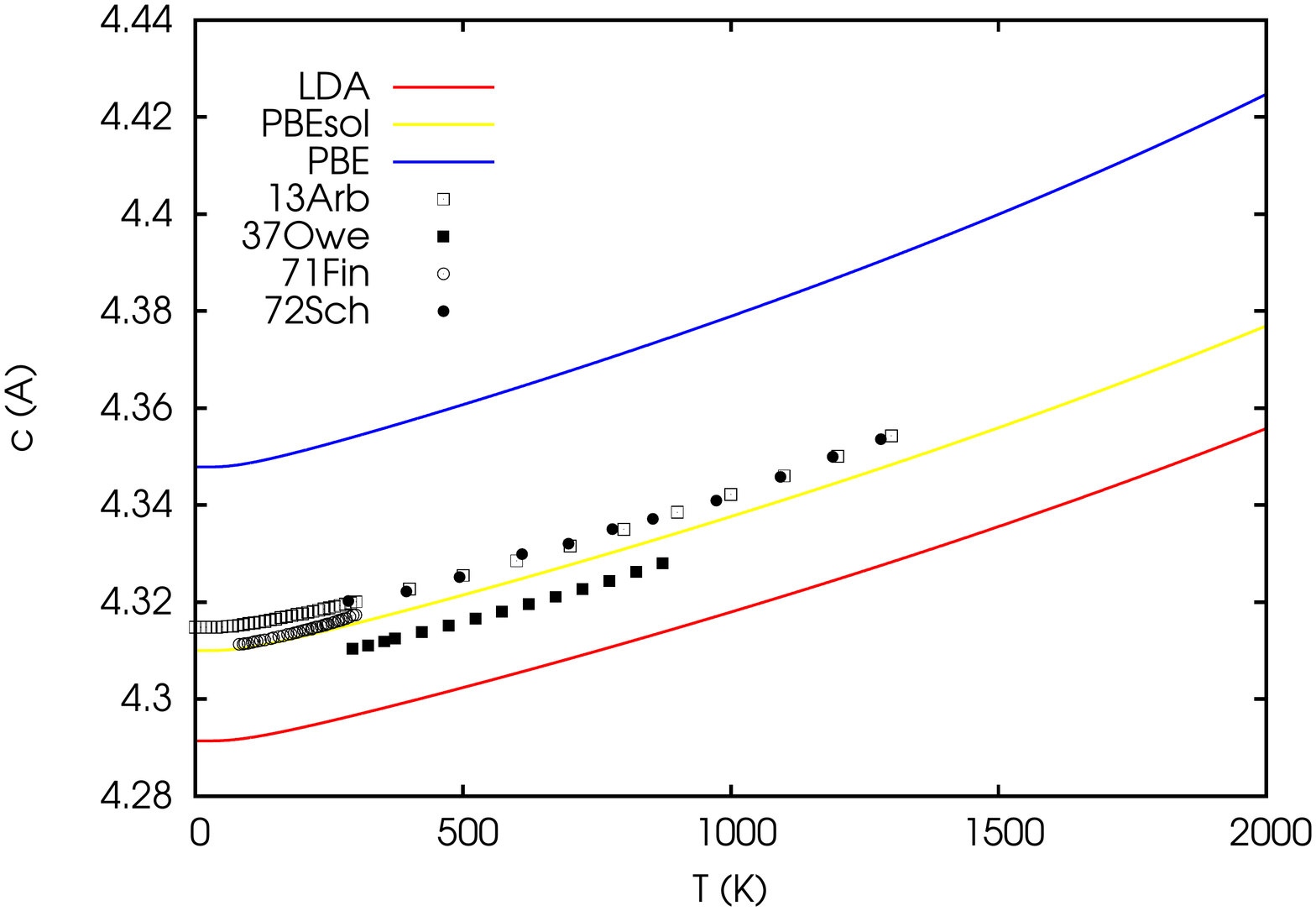}}
\includegraphics[height=6.5cm]{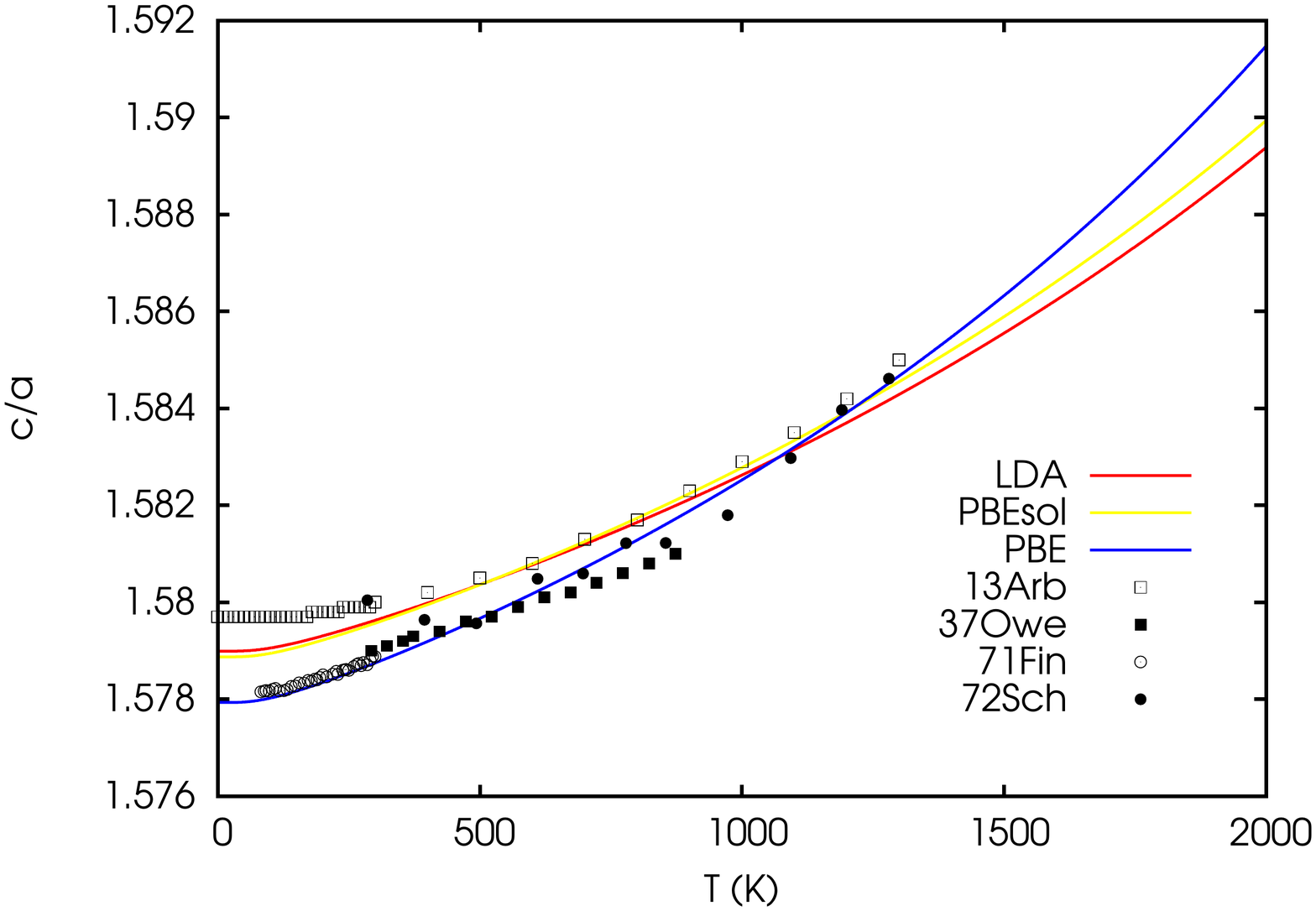}
\caption{Quasi-harmonic variation with temperature of the lattice parameters $a$, $c$ and $c/a$ of h.c.p. Os with both LDA and PBE functionals. 
These results were obtained using the full grid ($a$,$c/a$) and
minimizing the Helmholtz energy to obtain $a(T)$, $c(T)$ and $c/a(T)$.
Experimental data from different sources are shown as points: 37Owe~\cite{37Owe}, 71Fin~\cite{71Fin}, 72Sch~\cite{72Sch}. Points from 13Arb~\cite{13Arb} 
are not pure experimental data but assessed values obtained from a critical evaluation of available experimental data.
}\label{fg:Oslat}
\end{figure}

Next, we computed anisotropic properties using the full ($a$, $c/a$) grid. The temperature dependence of the lattice parameters obtained in this way
is reported in Fig.~\ref{fg:Oslat}. As found for other systems~\cite{paperReTc,14Pal}, the temperature dependence of the quasi-harmonic results
is in good agreement with the experiments, while the discrepancy between the calculated and experimental values
can be mostly ascribed to the exchange-correlation functional. In fact, LDA and PBE results essentially ``bracket'' the experimental values in the temperature range
of the figure. The experimental values from Owen et al.~\cite{37Owe} are lower than other data sets, but within the difference between LDA/PBE 
calculated results. We remark that the PBEsol results are
in remarkable good agreement with the experimental lattice parameters and within the experimental scatter.
The behavior of the $c/a$ ratio is similar to other h.c.p. elements~\cite{paperReTc}, i.e. its variation with temperature is very small ($<$1\% from 0 to 2000~K
in the LDA results) and compares satisfactorily with the experiments. No previously calculated result is available.

\begin{figure}[htp]
\hbox{\includegraphics[height=6.5cm]{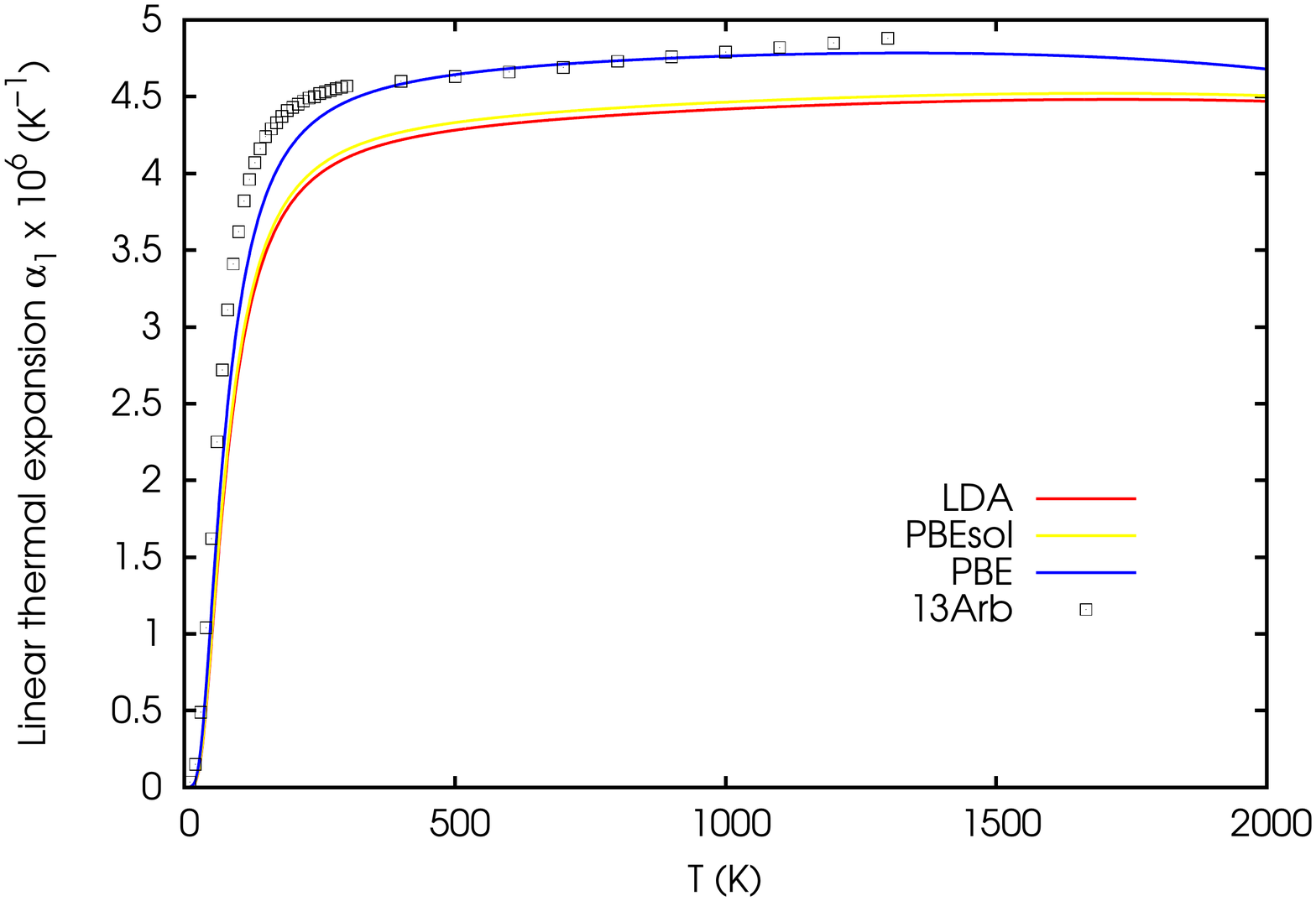}
\includegraphics[height=6.5cm]{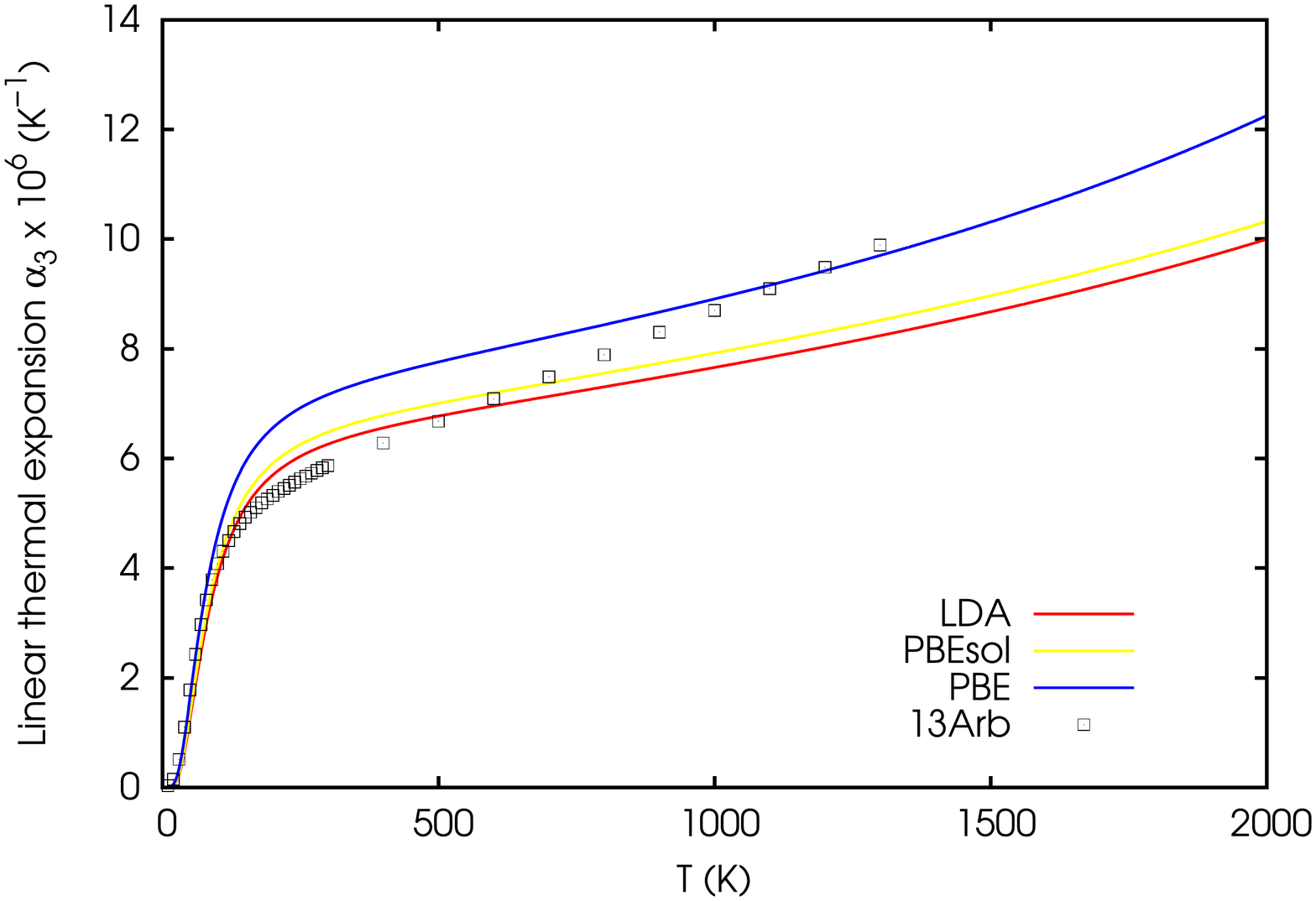}}
\caption{Calculated linear thermal tensor ($\alpha_1=\alpha_2$ and $\alpha_3$) for Os h.c.p. These results were obtained using the full grid ($a$,$c/a$),
minimizing the Helmholtz energy to obtain $a(T)$ and $c/a(T)$. Points from 13Arb (ref.~\cite{13Arb}) 
are not pure experimental data but assessed values obtained from a critical evaluation of available experimental data.}
\label{fg:Os_alphas}
\end{figure}

The results for the thermal expansion tensor are shown in Fig.~\ref{fg:Os_alphas}. We report only the ``assessed'' values from Arblaster~\cite{91Arb} 
for comparison. The values are obtained from a polynomial fit of the original experimental data of lattice parameters
(shown in Fig.~\ref{fg:Oslat}). The calculated results appear to be somehow controversial. The PBE results for $\alpha_1$ are in satisfactory 
agreement with Arblaster's assessed values, while LDA values underestimate them. On the contrary, both LDA and PBE calculated results for $\alpha_3$ do not 
reproduce well Arblaster's assessed values.

\subsection{Ruthenium}

The calculated lattice parameters of h.c.p. Ru are first compared with experimental the theoretical results from different sources in Tab.~\ref{tab:Rulatticeparameters}.
The experimental data have been taken from Ref.~\cite{13Arb2}, where the lattice parameters values have been corrected to 293~K using thermal expansion data.
The PBEsol results are in better agreement with the experiments than the LDA and PBE values. Contrary to Os, however, the PBEsol values are slightly
lower than the experimental values. Theoretical results at 0~K from Heid et al.~\cite{00Hei} using an LDA functional and a norm-conserving Hamann-Sch\"uter-Chiang
pseudopotential appear to be significantly lower than our values, possibly because of the different pseudopotential constructed using a mixed-basis set. 
The results from Yaozhuang et al., also using a norm-conserving pseudopotential and an LDA functional, appear to be closer to ours for the lattice parameter
$a$, but significantly differ for $c$. Their results for the latter parameter (and for the $c/a$ ratio) are also quite different from all experimental results.
Finally, the results from Souvatzis et al. were obtained using a PAW pseudopotential and PBE functional. As can be seen
from Tab.~\ref{tab:Rulatticeparameters}, they overestimate both our results (PBEsol and PBE) and all experimental results. 

\begin{table}[htp]
\caption{\label{tab:Rulatticeparameters} Comparison between structural parameters of h.c.p. Ru from different sources.  
Experimental data are taken from Ref.~\cite{13Arb2} and references therein. They have been corrected to 293 K using thermal expansion data. 
Previous theoretical results are also reported for comparison.}
\begin{tabular}{@{}lcccc}
 Method & $a$ & $c$ & $c/a$ \\
        & $\AA$ & $\AA$ &  \\
PAW, LDA (0~K, this work) & 2.6770 & 4.2260 & 1.5786  \\
PAW, PBEsol (0~K, this work) & 2.6920 & 4.2485 & 1.5784  \\
PAW, PBE (0~K, this work) & 2.7219 & 4.2934 & 1.5774  \\
PAW, LDA (293~K, this work) & 2.6820 & 4.2363 & 1.5795 \\
PAW, PBEsol (293~K, this work) & 2.6967 & 4.2598 & 1.5796 \\
PAW, PBE (293~K, this work) & 2.7273 & 4.3056 & 1.5787 \\
NC, LDA (0~K, Heid et al.) &  2.701 & 4.270  & 1.581  \\
PAW, GGA (0~K, Souvatzis et al.) & 2.74 & 4.32 & 1.577 \\
NC, LDA (0~K, Yaozhuang et al.) & 2.691 & 4.322 & 1.606 \\ 
Exp. (Owen et al. 1935)    & 2.7044 & 4.2818 & 1.5833 \\
Exp. (Owen and Roberts 1936)    & 2.7042 & 4.2819 & 1.5834 \\
Exp. (Owen and Roberts 1937)    & 2.7040 & 4.2819 & 1.5835 \\
Exp. (Ross et al.)    & 2.7042 & 4.2799 & 1.5827 \\
Exp. (Finkel et al. 1971)    & 2.7062 & 4.2815 & 1.5821 \\
Exp. (Hellawell et al. 1954)    & 2.7058 & 4.2817 & 1.5824 \\
Exp. (Swanson et al. 1955)    & 2.7059 & 4.2819 & 1.5824 \\
Exp. (Hall et al. 1957)    & 2.7058 & 4.2805 & 1.5820 \\
Exp. (Anderson et al. 1960)    & 2.7058 & 4.2814 & 1.5823 \\
Exp. (Cernohorsky 1960)    & 2.7059 & 4.2812 & 1.5822 \\
Exp. (Savitskii et al. 1962)    & 2.7059 & 4.2819 & 1.5824 \\
Exp. (Schroeder et al. 1972)    & 2.7056 & 4.2826 & 1.5829 \\
\end{tabular}
\end{table}

\begin{figure}[htp]
\hbox{\includegraphics[height=6.5cm]{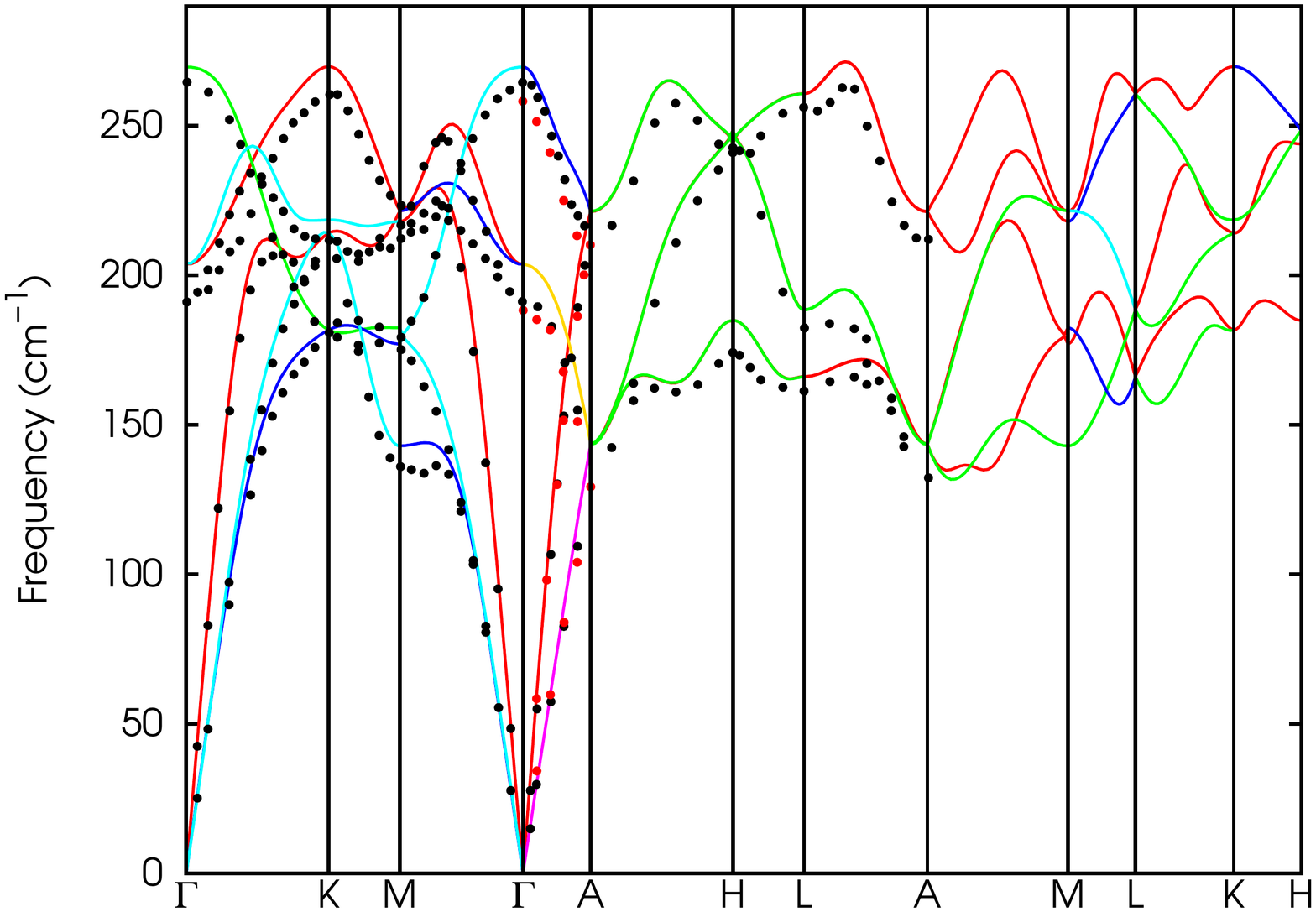}
\includegraphics[height=6.5cm]{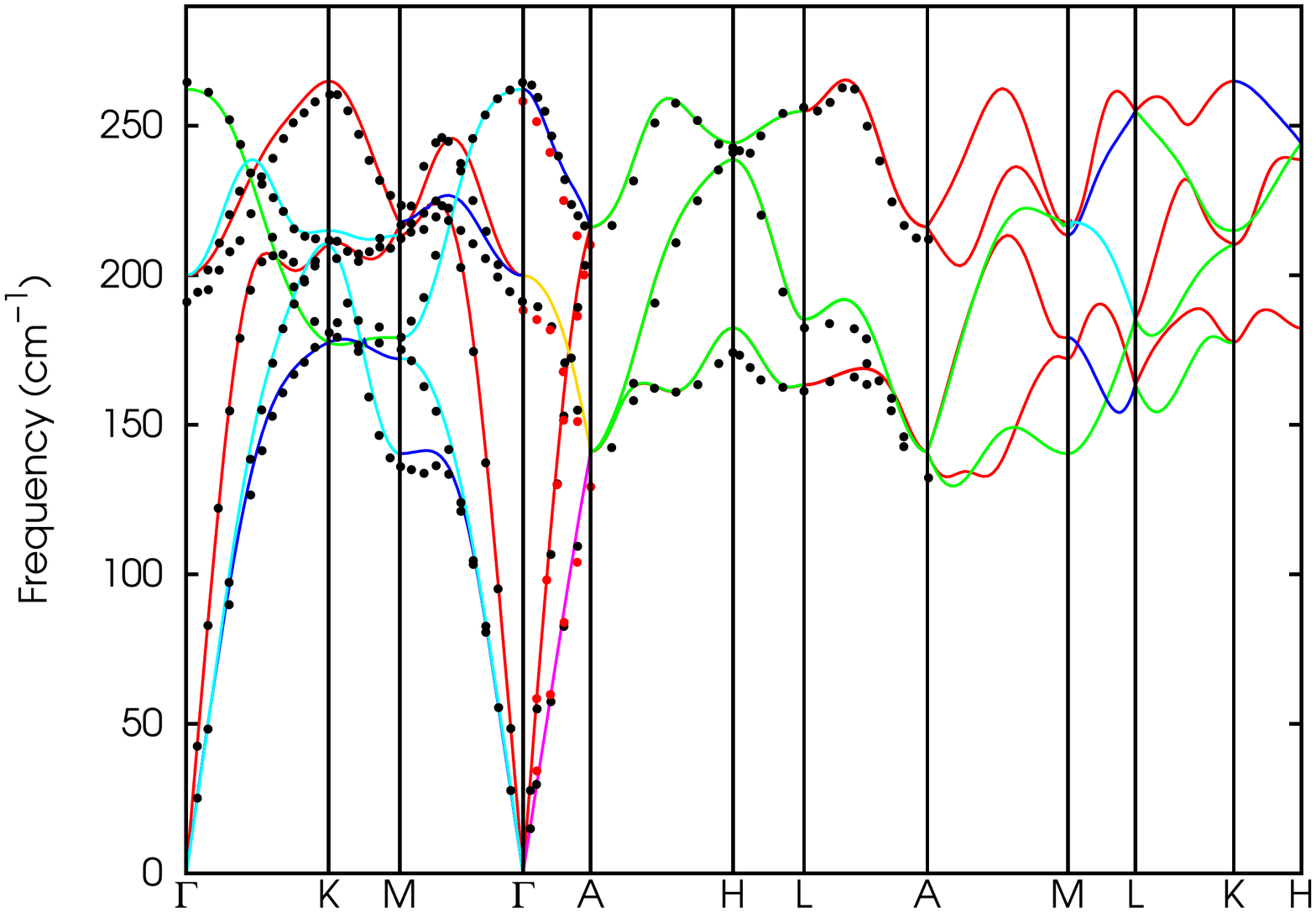}}
\includegraphics[height=6.5cm]{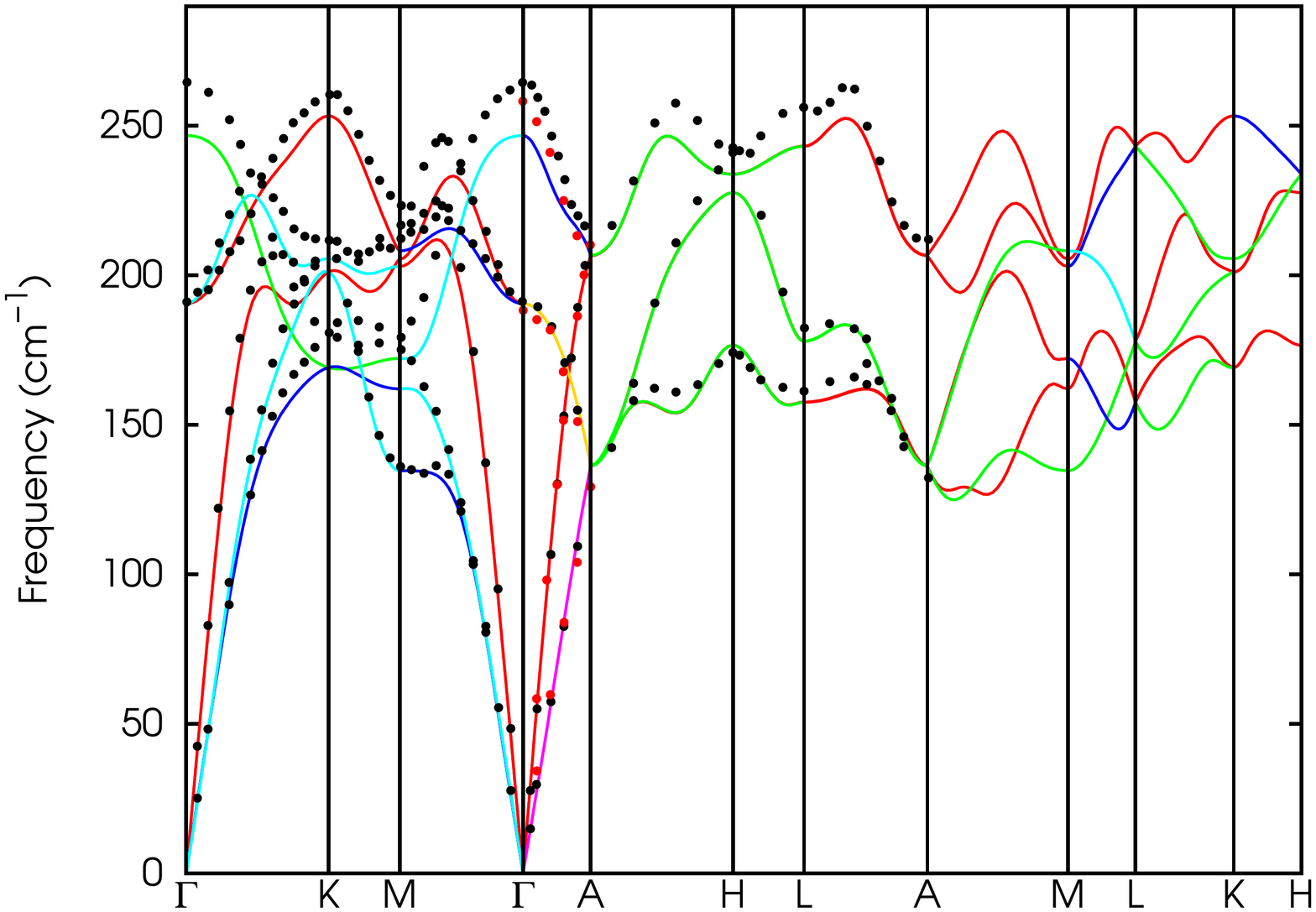}
\caption{Phonon dispersion curves of Ru along high symmetry points and directions in the BZ calculated with
LDA (a), PBEsol (b) and PBE (c) exchange-correlation functionals using a mesh of 6$\times$6$\times$4 q-points. 
The colours represent the different simmetries of the phonon branches (for details see the thermo\_pw documentation).
The calculation was carried out at the equilibrium lattice parameters at 298~K 
($a=2.6820$ \AA~and $c/a=1.5795$ for LDA, $a=2.6967$ \AA~and $c/a=1.5796$ for PBEsol, $a=2.7273$ \AA~and $c/a=1.5787$ for PBE). Experimental data are from
Ref.~\cite{00Hei} (black circles) and Ref.~\cite{81Smi} (red circles).
}\label{fg:Rudisp}
\end{figure}

\begin{figure}[htp]
\hbox{\includegraphics[height=7cm]{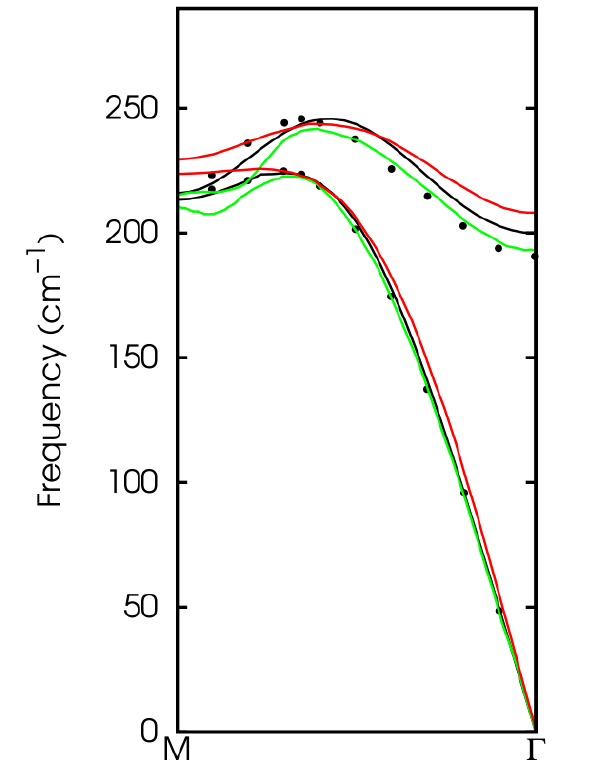}
\label{fg:RudispGM1}
\includegraphics[height=7cm]{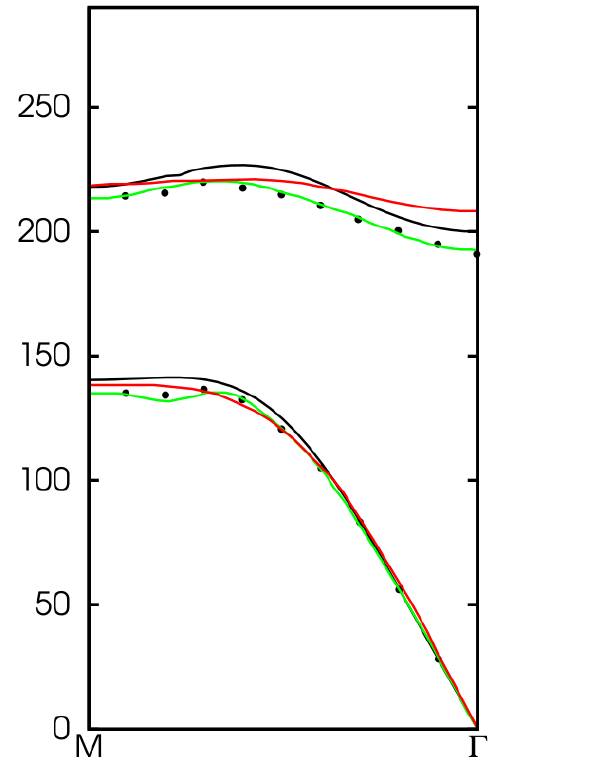}
\label{fg:RudispGM2}
\includegraphics[height=7cm]{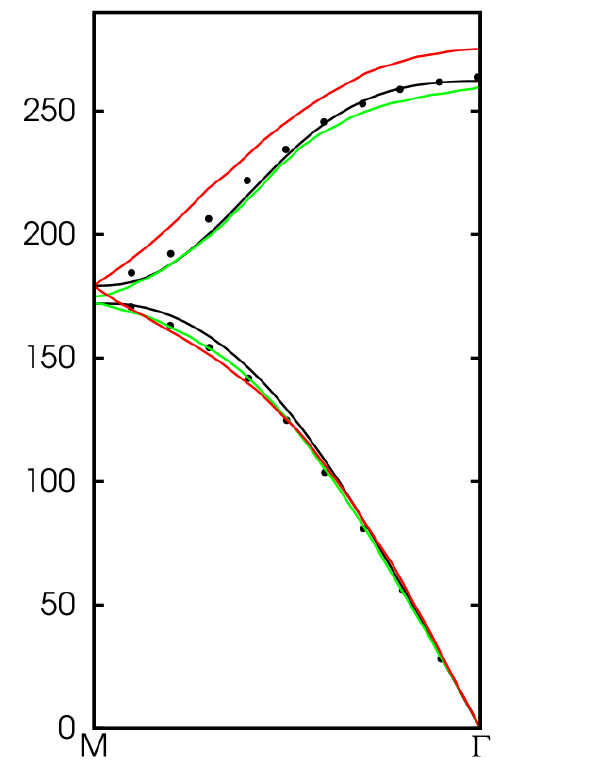}
\label{fg:RudispGM3}}
\caption{Phonon dispersion curves of Ru along the $\Gamma$-M direction in the BZ calculated as in this work at 298~K using DFPT and the 
PBEsol functional (black lines),
 as in Heid et al.~\cite{00Hei} at 0~K using DFPT and an LDA functional (yellow lines) and as in Souvatzis et al.~\cite{08Sou} at 0~K using a 
 the supercell method and a GGA functional (red lines).  
Experimental data are from
Ref.~\cite{00Hei} (black circles). (a) longitudinal modes $\Sigma_1$, (b) transversal modes $\Sigma_4$, 
(c) transversal modes $\Sigma_3$.
}\label{fg:RudispGM}
\end{figure}

Contrary to Os, experimental values of the phonon frequencies are available for h.c.p. Ru and are shown in Fig.~\ref{fg:Rudisp} together 
with the theoretical results calculated using LDA, PBEsol and PBE
functionals. The LDA frequencies overestimate the experimental values in most branches, up to nearly 20 cm$^{-1}$. 
The PBE results are in better agreement with the experimental frequencies, although a slight difference is still present 
at the high frequency optical branches where they underestimate the experiments.
PBEsol frequencies agree fairly well with the experiments for almost every branch. This functional dependence of the phonon
frequencies is similar to what was found in several cubic metals~\cite{13Dal}.
The phonon dispersions in Fig.~\ref{fg:Rudisp}
also agree well with the results calculated by Heid et al.~\cite{00Hei} using density-functional-perturbation theory and 
by Souvatzis et al.~\cite{08Sou} using a 4$\times$4$\times$3 supercell. As an example, we report in Fig.~\ref{fg:RudispGM}
a comparison of previously calculated frequencies and experimental data along the $\Gamma$-M direction in the BZ, where some
anomalies with respect to other h.c.p. elements have been found in the vicinity of the M point: (1) the longitudinal
modes (Fig.~\ref{fg:RudispGM1}) show a sudden drop in frequency when approaching the M point; (2) the acoustic branch of the transversal
modes $\Sigma_4$ (Fig.~\ref{fg:RudispGM2}) are essentially flat for about 1/3 of the $\Gamma$-M distance; (3) the z-polarized branches $\Sigma_3$
are nearly degenerate at the M point.
All points are well reproduced by all calculated results. The agreement with the experimental data is also remarkably good for all calculated
frequencies, with Souvatzis' calculated frequencies which appear slightly worse than Heidi's and ours, despite the fact that different functionals were used
and some values were obtained at 0~K. The fact can possibly be explained noting that Heidi's calculated LDA lattice parameters at 0~K are very close
to our PBEsol calculated values at 298~K.

\begin{figure}[htp]
\includegraphics[height=10cm]{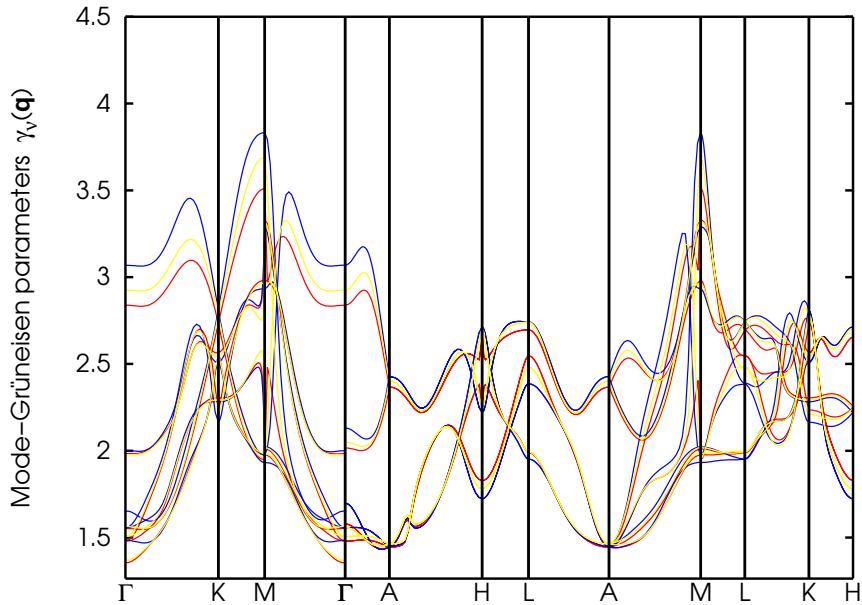}
\caption{Volume-dependent mode Gr\"{u}neisein parameters of h.c.p. Ru calculated with LDA (red line), PBEsol (yellow line) and PBE (blue line)
exchange-correlations functional at the equilibrium 
lattice parameters at 300~K.  
}\label{fg:Rugrun}
\end{figure}

The Gr\"uneisen parameters calculated using the LDA, PBEsol and PBE functionals are shown in Fig.~\ref{fg:Rugrun}. They are remarkably similar to those of Os, with 
the highest values at the M point and ranging from slightly less than 1.5 to approximately 3.5.


\begin{figure}[htp]
\hbox{
\includegraphics[height=6cm]{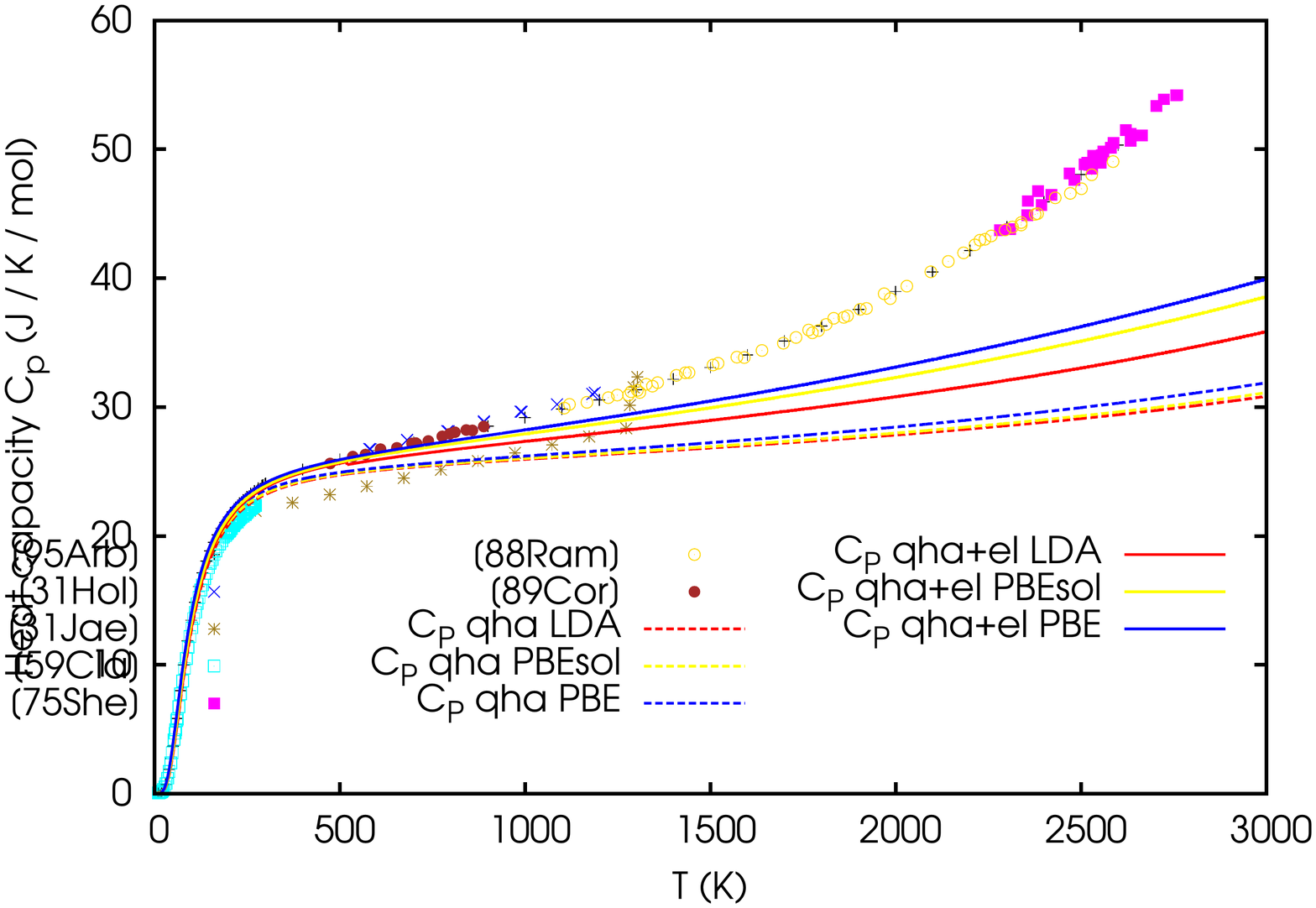}
\label{fg:RuCp1}
\includegraphics[height=6cm]{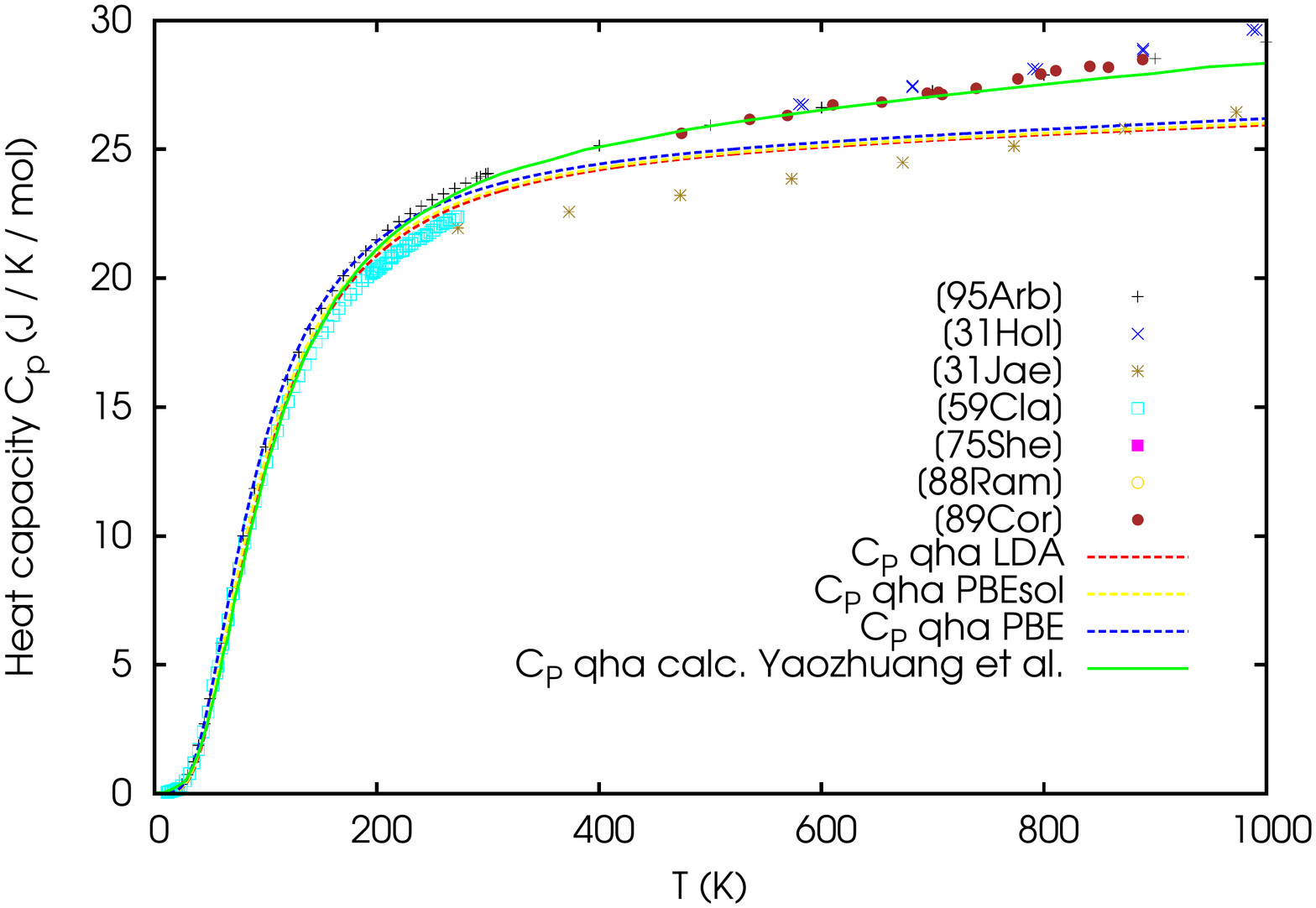}
\label{fg:RuCp2}}
\caption{(a) Calculated isobaric heat capacity of h.c.p. Ru as a function of temperature with LDA/PBEsol/PBE exchange-correlation functionals, 
quasi-harmonic only ($C_P$ qha) and 
quasi-harmonic plus electronic ($C_P$ qha+el). The calculations were done assuming a constant $c/a$ ratio and using the Murnaghan EOS. 
Experimental data from different sources are shown as points: 31Hol~\cite{31Hol}, 31Jae~\cite{31Jae}, 59Clu~\cite{59Clu}, 75She~\cite{75She},
88Ram~\cite{88Ram} and 89Cor~\cite{89Cor}. Points from Arblaster (Ref.~\cite{95Arb}) 
are not pure experimental data but assessed values obtained from a critical evaluation of available experimental data; (b) as in (a) but including
previous theoretical results from Yaozhuang et al. (and removing qha+el results for clarity).
}\label{fg:RuCp}
\end{figure}

\begin{figure}[htp]
\includegraphics[width=13cm]{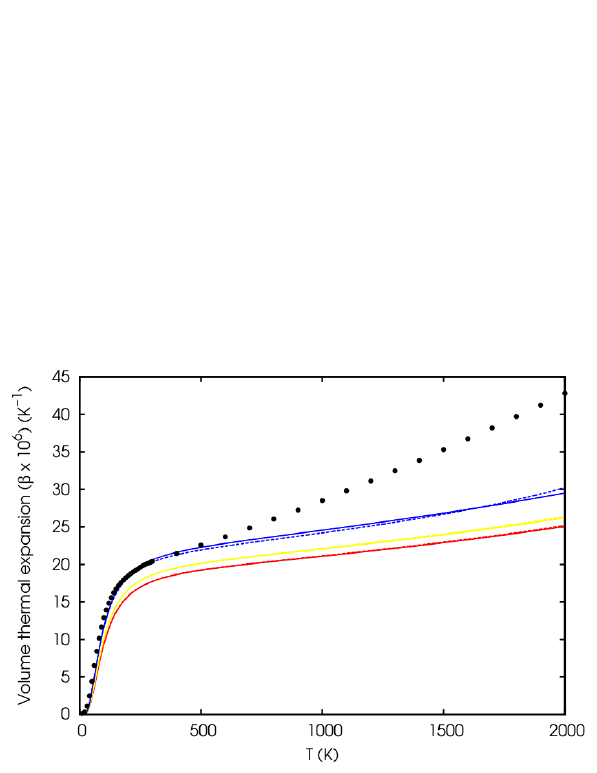}
\caption{Quasi-harmonic volume thermal expansion of h.c.p. Ru calculated as a function of temperature with LDA (red lines), PBEsol (yellow line)
and PBE (blue lines) exchange-correlation functionals.
The continuous lines are obtained
assuming a constant $c/a$ ratio and using the Murnaghan EOS. The dashed lines are obtained using the full anisotropic results for the linear thermal 
expansions ($\alpha_1$ and $\alpha_3$) and $\beta=2 \alpha_1 + \alpha_3$. Theoretical results from Yaozhuang et al.~\cite{07Yao} (olive line)
are also reported for comparison.
}\label{fg:Rubetas}
\end{figure}

As for Os, we first carried out calculations of isotropic quantities of h.c.p. Ru using Murnaghan EOS and a constant $c/a$ ratio. The results for
the isobaric heat capacity, for which several experimental data are available, are shown in Fig.~\ref{fg:RuCp1}. It must first be remarked that the experimental
values from Jaeger et al.~\cite{31Jae} are significantly lower compared to all other data in the temperature range 500-1500~K. Furthermore, the low temperature
data from Clusius~\cite{59Clu} show a trend at high temperatures which conflicts with other experimental values above 300~K. In fact, Arblaster discussed
this issue in his work and his ``assessed'' values (also reported in the figure) have been chosen to properly link low temperature data 
from Clusius~\cite{59Clu} to high temperature data. It is remarkable that our first-principles PBE results are nearly identical to Arblaster's values
up to 500~K, thus confirming his choice of recommended values. Above this temperature, both LDA and PBE calculated results start deviating from the experimental
points and at high temperature the calculations largely underestimate the experiments.
It is noteworthy that the electronic contribution to the heat capacity of Ru is not negligible and necessary, though not sufficient, to improve the agreement with the 
experimental data. It is also worth noting that there is a significant difference in this contribution for h.c.p. Ru 
between the LDA and PBE results. This arises from the shift of the electronic DOS computed at different (minimum) lattice parameters when 
the calculation is done with different functionals.
Fig.~\ref{fg:RuCp2} shows an enlargement of the isobaric heat capacity results in the temperature range 0-1000~K, including previously calculated results by 
Yaozhuang et al. (for the sake of clarity, our results including the electronic contribution have been removed). It is clear that our heat capacity values 
agree well with those computed by Yaozhuang et al. at low temperatures, whereas above approximately room temperature the latters are higher and closer to the experimental
data without including the electronic contribution. These authors state in their paper that they computed the heat capacity for the anisotropic h.c.p. Ru
using Eq.~\ref{eq:Cp} as in our present work. In this equation, they used the volumetric thermal expansion obtained 
from anisotropic linear thermal expansions as $\beta=2 \alpha_1 + \alpha_3$. As we prove in the following (Fig.~\ref{fg:Rubetas}), the difference 
with the volumetric thermal expansion obtained in our calculations is negligible. Negligible differences are expected also in the 
values of $C_V$, which at room temperature have almost reached the plateau value of $3R$, with $R$ the gas constant. 
Yaozhuang et al. used in Eq.~\ref{eq:Cp} the anisotropic bulk modulus obtained from the full elastic constants tensor. They report
a temperature-independent value of the bulk modulus, derived from elastic constants computed only at 0~K, equal to 322 GPa at 0~K and
which compares reasonably well with our results (361 and 307 GPa for LDA and PBE respectively using the Murnaghan
EOS and including ZPE). In our results the bulk modulus decreases with temperature down to 325 GPa and 271 GPa for LDA and PBE respectively at 1000~K.
Hence this may explain the discrepancy between Yaozhuang's and our results for the high temperature isobaric heat capacity, with an additional contribution
from differences in the equilibrium volumes, stemming from 
discrepancies in the calculated lattice parameters (as in Tab.~\ref{tab:Rulatticeparameters}). In fact the volume obtained from the 0~K
parameters from Yaozhuang et al. is 27.1 $\AA^3$ and ours, including the ZPE, are 26.3 (LDA) and 27.6 (PBE) $\AA^3$. 
Thus the excellent agreement with the experimental data obtained in Yaozhuang's quasi-harmonic results without electronic contribution is likely fortuitous.

\begin{figure}[htp]
\hbox{
\includegraphics[height=6.5cm]{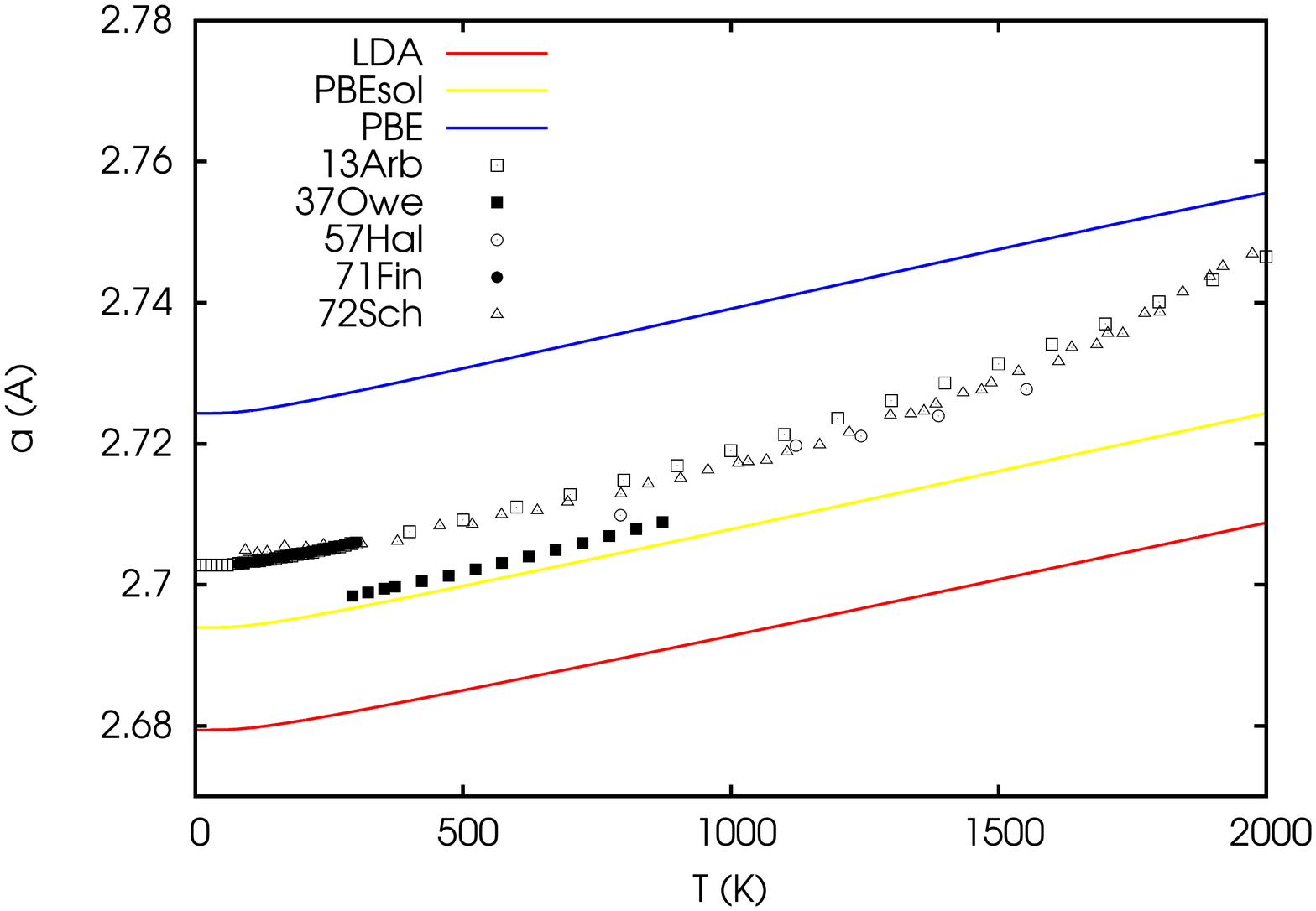}
\includegraphics[height=6.5cm]{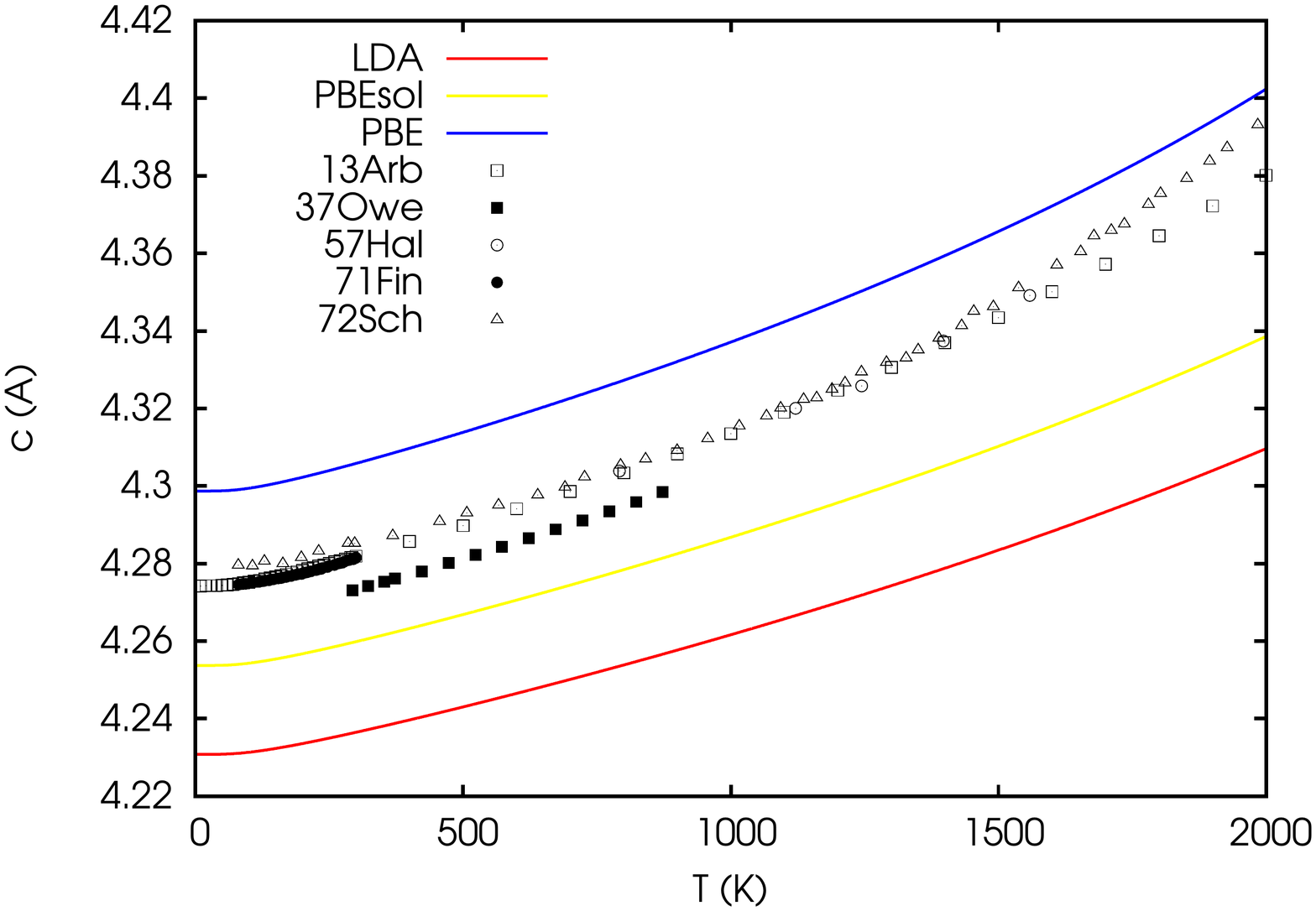}}
\includegraphics[height=6.5cm]{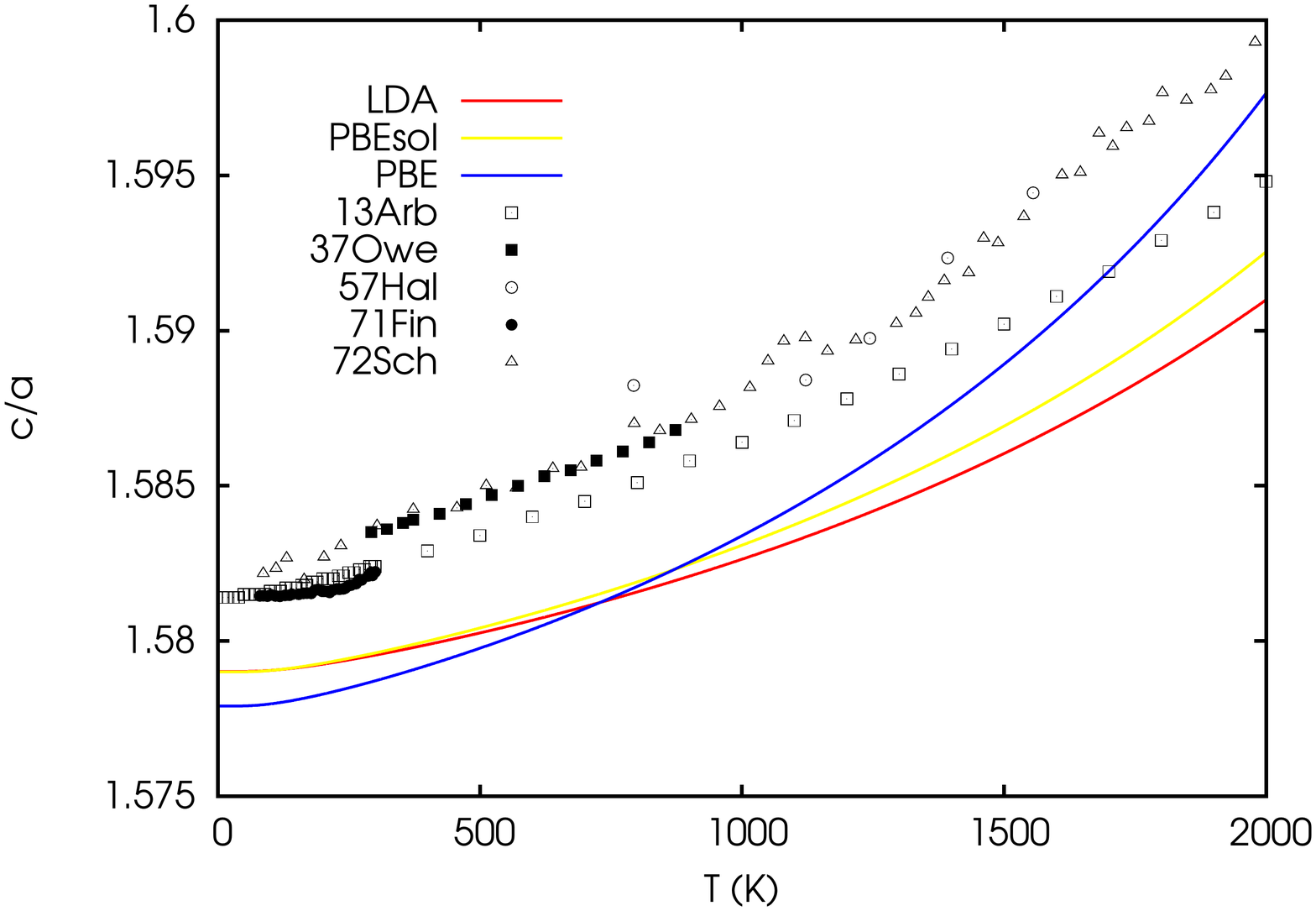}
\caption{Quasi-harmonic variation of the lattice parameters $a$, $c$ and $c/a$ of h.c.p. Ru. 
Experimental data from different sources are shown as points: 37Owe~\cite{37Owe}, 57Hal~\cite{57Hal}, 71Fin~\cite{71Fin}, 72Sch~\cite{72Sch}.
Points from 13Arb~\cite{13Arb2} 
are not pure experimental data but assessed values obtained from a critical evaluation of available experimental data.
These results were obtained using the full grid ($a$,$c/a$),
minimizing the Helmholtz energy to obtain $a(T)$, $c(T)$  and $c/a(T)$
}\label{fg:Rulat}
\end{figure}

The lattice parameters of h.c.p. Ru as a function of temperature calculated using the full ($a$, $c/a$) grid are reported in Fig.~\ref{fg:Rulat}. 
As for Os, the LDA and PBE results ``bracket'' the experimental values starting from 0~K, with a difference which is due to the exchange-correlation
functional. In fact, the PBEsol results are closer to the experiments, but contrary to Os they are still slightly underestimating all data sets.
The temperature dependence of the experimental lattice parameters is well reproduced by the quasi-harmonic results up to approximately
1500~K. At higher temperatures the quasi-harmonic approximation breaks down and the experimental values increase more than our results.
The variation of the $c/a$ ratio for Ru is slightly higher than for Os ($\sim1\%$) but still very small. The calculated results do not
reproduce well the experimental values, though the differences are of the same order of magnitude as those among different experimental data sets. The different
temperature dependence of the $c/a$ ratio is probably within the numerical accuracy which can be expected from these calculations.

\begin{figure}[htp]
\hbox{\includegraphics[height=6.5cm]{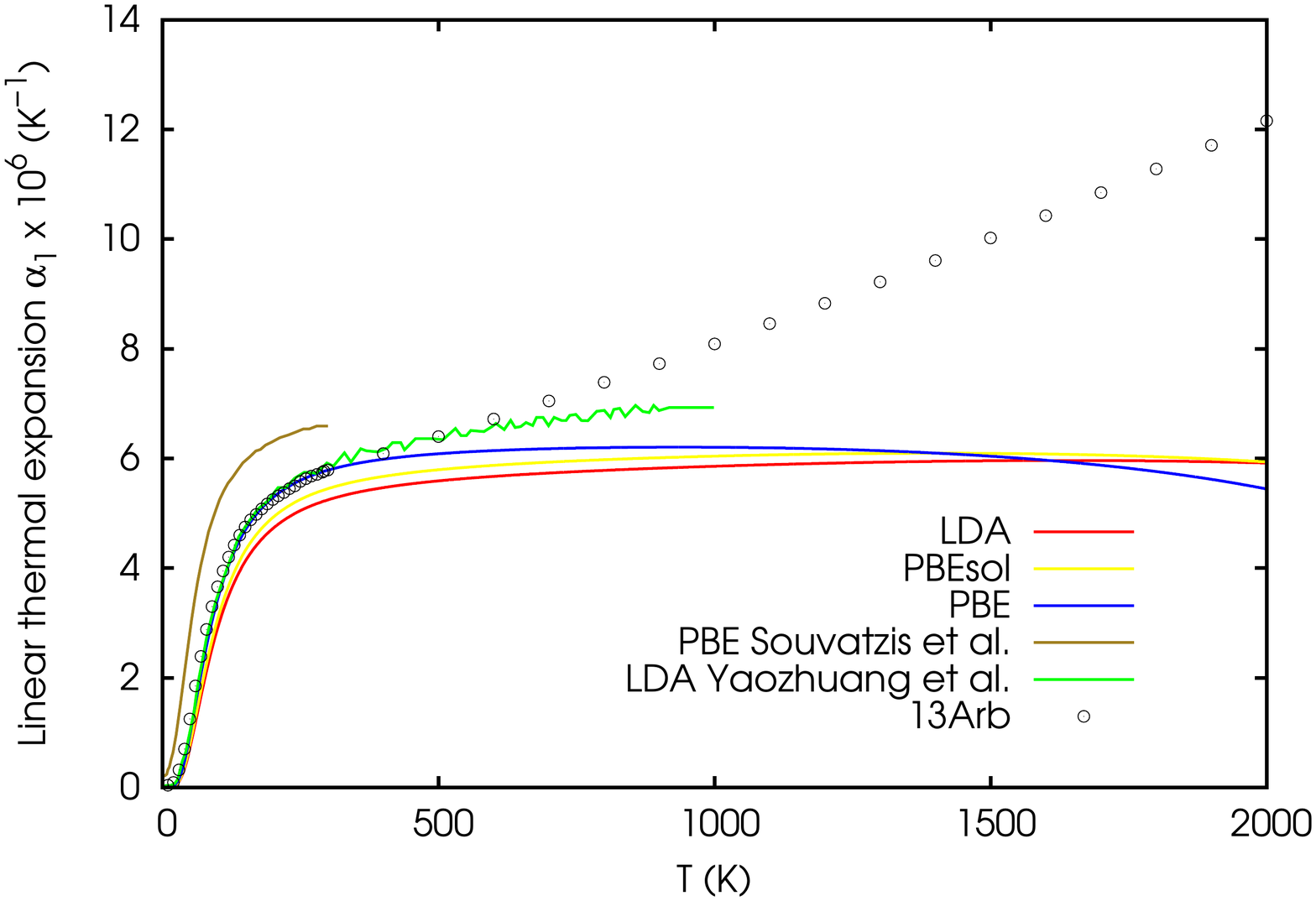}
\includegraphics[height=6.5cm]{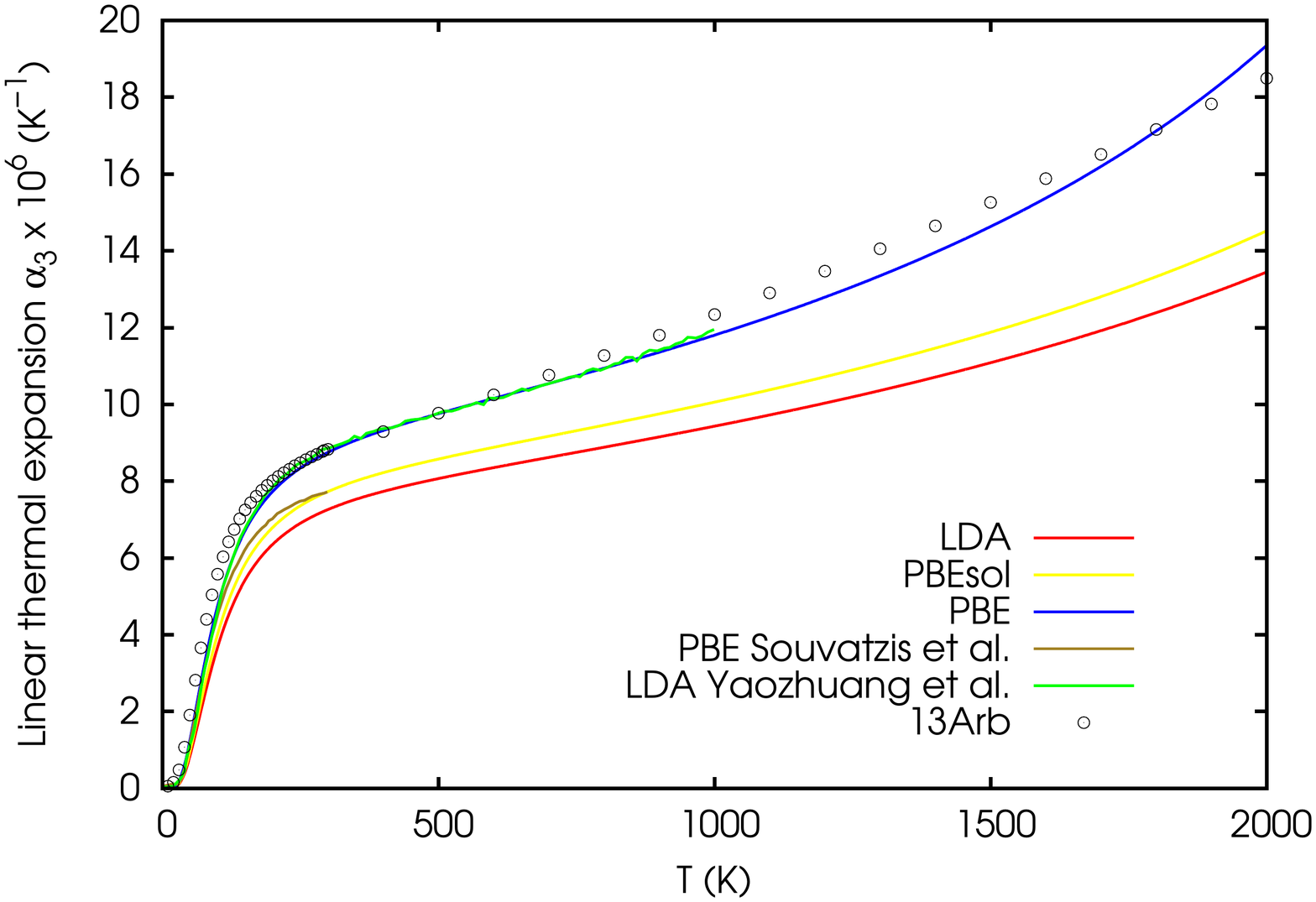}}
\caption{Calculated linear thermal expansion tensor ($\alpha_1=\alpha_2$ and $\alpha_3$) for Ru h.c.p. using LDA, PBE and PBEsol functionals.
These results were obtained using the full grid ($a$,$c/a$),
minimizing the Helmholtz energy to obtain $a(T)$ and $c/a(T)$. Points from 13Arb~\cite{13Arb2} 
are not pure experimental data but assessed values obtained from a critical evaluation of available experimental data. Previously calculated results
from Souvatzis et al.~\cite{08Sou} (olive line) and Yaozhuang et al.~\cite{07Yao} (green line) are also reported for comparison.}
\label{fg:Ru_alphas}
\end{figure}

The anisotropic thermal expansions for h.c.p. Ru are shown in Fig.~\ref{fg:Ru_alphas}. At low temperature, a good agreement is obtained between the
experimental values and the PBE results, whereas the LDA calculated values are lower in the whole temperature range up to 2000~K. At high temperatures, the calculated thermal expansion
along the $a$ direction ($\alpha_1$) strongly underestimate the experimental values, while the good agreement between the PBE results and the experiments for 
$\alpha_3$ is probably fortuitous. In fact, results for $\alpha_1$, the lattice parameters and the heat capacity all suggest that the quasi-harmonic 
approximation breaks down for Ru above an approximate temperature which depends on the evaluated quantity and goes from 500~K for the heat capacity
(Fig.~\ref{fg:RuCp}) to 1000~K for the lattice parameters (Fig.~\ref{fg:Rulat}). Our present values of $\alpha_1$ and $\alpha_3$ agree almost 
perfectly with the previously calculated values by Yaozhuang et al.~\cite{08Sou}, while they differ significantly from those by Souvatzis et al., which are
also limited to temperatures below 300~K. 

\begin{figure}[htp]
\includegraphics[width=13cm]{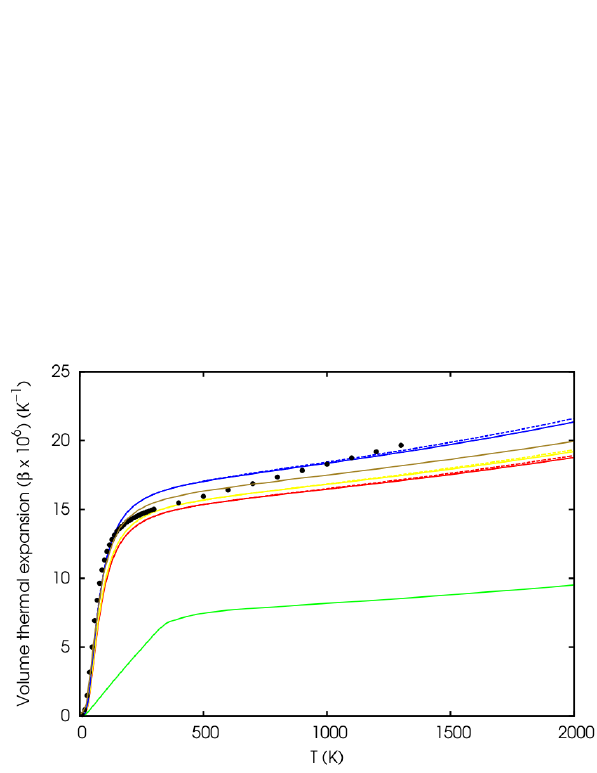}
\caption{Quasi-harmonic volume thermal expansion of h.c.p. Os calculated as a function of temperature with LDA (red lines), PBEsol (yellow lines)
and PBE (blue lines) exchange-correlation functionals. 
The continuous lines are obtained
assuming a constant $c/a$ ratio and using the Murnaghan EOS. The dashed lines are obtained using the full anisotropic results for the linear thermal 
expansions ($\alpha_1$ and $\alpha_3$) and $\beta=2 \alpha_1 + \alpha_3$. Theoretical results from Deng et al.~\cite{11Den} (olive line)
and Liu et al.~\cite{11Liu} (green line) are also reported for comparison.
}\label{fg:Osbetas}
\end{figure}

We finally point out that the calculation of isotropic quantities, such as volume thermal expansion, is nearly identical when using the Murnaghan EOS
and a constant $c/a$ ratio and when using the full ($a$, $c/a$) grid. This is shown in Figs.~\ref{fg:Osbetas},~\ref{fg:Rubetas}. As we already discussed, the variation
of the $c/a$ ratio with temperature is limited for both Os and Ru and hence has limited influence on isotropic quantities. Similar results were also obtained for
Re and Tc~\cite{paperReTc}. The thermal expansions for both Os and Ru are positive in the 
entire temperature range, in agreement with the calculated mode Gr\"uneisen parameters. The calculated thermal expansion of Os is also in good agreement with
the theoretical results by Deng et al.\cite{11Den}, as it can be seen in  Fig.~\ref{fg:Osbetas}. On the contrary, the quasi-harmonic Debye model used
in the work by Liu et al. appears to significantly underestimate the volume thermal expansion compared to both full quasi-harmonic results and
Arblaster's assessed values.

At the end of our discussion of several results for different quantities, 
we note that the quasi-harmonic approximation is expected to break down at high temperatures. This is well 
reported in several papers (and references therein)~\cite{14Pal, 07Gra, 10Bar}.
It is customary to set the Debye temperature ($\theta_D$) as the breaking point, above which anharmonic contributions 
beyond the quasi-harmonic approximation start
playing a non-negligible role, more recent works suggest that the breaking point is rather system dependent and there are cases in which quasi-harmonic 
results provide satisfactory values well above $\theta_D$. Methods to include additional anharmonic contributions are mostly based on different 
flavors of molecular dynamics or on including third- or higher-order terms in the perturbation expansion of the internal energy in 
atomic displacements~\cite{15Gle, 14Err, 14Pal}. These methods are more computational demanding, unless they include some simplifying assumptions.  
The quasi harmonic approximation
still provide reliable results at low/medium temperatures and often represents a starting point for the application of more sophisticated methodologies.
A final note on the high-pressure anomalies recently found in h.c.p. Os by Dubrovinsky et al.~\cite{15Dub}. As well discussed in their work, these anomalies
would require a higher level of theory (DMFT, spin-orbit coupling). The present quasi-harmonic calculations however use phonons calculated at volumes 
corresponding to the maximum pressure of about 90~GPa. Up to this point, the difference between the DMFT results and DFT calculations appear to be negligible.

\section{Conclusions}
Using density functional theory and the quasi-harmonic approximation, we analyzed the lattice dynamics and thermophysical properties 
of h.c.p. Os and Ru. At first, we employed a temperature-independent $c/a$ and the Murnaghan EOS to derive
temperature-dependent properties. Afterwards, we carried out quasi-harmonic calculations on a full ($a$, $c/a$) grid and obtained fully 
anisotropic results as the temperature dependence of the lattice parameters and thermal expansion tensor.

Both Os and Ru phonon dispersions do not show soft modes as we obtained for Re and Tc. A good agreement with the experimental 
phonon frequencies was obtained for h.c.p. Ru. A slight disagreement with
experiments remains for Ru for the optical high-frequency modes. A satisfactory agreement with the experiments 
was also obtained for the TO pressure shift at the 
$\Gamma$ point for Os, particularly when using the PBEsol functional.

For Os, the calculated thermophysical quantities show a remarkable agreement with experimental values up to the maximum temperature of 1300~K where they are available.
The lattice parameters as a function of temperature show a difference with experimental values at any temperature which mostly depends on the exchange-correlation 
functional, with LDA (PBE) calculations underestimating (overestimating) the experiments, whereas the PBEsol values are essentially on top of experiments. 
The PBE thermal expansion along the $a$ direction is in satisfactory
agreement with the assessed values from Arblaster, while along the $c$ direction no functional is fully in agreement with the recommended values.

For Ru, experimental data of thermodynamic quantities are available up to nearly 3000~K and the quasi-harmonic approximation appears to break down between 500-1000~K,
depending on the quantity considered. For example, the calculated values of the heat capacity compare well with experiments up to approximately 1000~K, whereas the 
temperature dependence of the calculated lattice parameters starts deviating from the experimental one above 1500~K. 
Including the electronic contribution to the heat capacity in Ru (and Os) is important, as is usually the case for metals.
The thermal expansion $\alpha_1$ is in good agreement with Arblaster's recommended values below 500~K, but at higher temperatures the difference is significant. On the contrary,
the remarkable agreement between the calculated thermal expansion $\alpha_3$ and Arblaster's values up to 2000~K is probably 
fortuitous since it does not correspond to a similar 
good agreement for $c$ and $c/a$. 

It was already pointed out for Re and Tc and confirmed here for Os and Ru that the temperature variation of the $c/a$ ratio obtained from 
the quasi-harmonic approximation on a full ($a$, $c/a$) grid is limited to 1\% or less up to 2000~K. This explains why the volume thermal expansion
obtained from the full grid calculation is very similar to that obtained from the Murnaghan EOS and a constant $c/a$ with temperature.

\section*{Acknowledgement}
This work was performed within the MaX Center of Excellence, with support from the European Union
Horizon 2020 EINFRA program under grant agreement No 676598. Computational facilities have been provided by SISSA through its Linux
Cluster and ITCS.

\end{document}